\documentstyle[11pt,aaspp4,flushrt]{article}
\begin{document}
\newcommand{\m}{\tilde{m}}
\newcommand{\dmhat}{\hat{\delta m}}
\def\figcaption[#1]#2{\begin{figure}\plotone{#1}\caption{#2}\end{figure}}
%


\title{The Expulsion of Stellar Envelopes in Core-Collapse Supernovae}
\author{Christopher D. Matzner and Christopher F. McKee}
\affil{Departments of Physics and Astronomy \\601 Campbell Hall \\
University of California, Berkeley, CA 94720 \\ {\bf Submitted to ApJ: 5/5/98.}}
\authoremail{matzner@astron.berkeley.edu, cmckee@mckee.berkeley.edu}

\begin{abstract}
We examine the relation between presupernova stellar structure and the
distribution of ejecta in core-collapse supernovae of types Ib, Ic and
II, under the approximations of adiabatic, spherically symmetric
flow. We develop a simple yet accurate analytical formula for the
velocity of the initial forward shock that traverses the stellar
envelope. For material that does not later experience a strong reverse
shock, the entropy deposited by this forward shock persists into the
final, freely-expanding state. We demonstrate that the final density
distribution can be approximated with simple models for the final
pressure distribution, in a way that matches the results of
simulations. Our results indicate that the distribution of density and
radiation pressure in a star's ejecta depends on whether the outer
envelope is radiative or convective, and if convective, on the
composition structure of the star.

Our models are most accurate for the high-velocity ejecta cast away
from the periphery of a star. For stellar structures that limit to a
common form in this region, the resulting ejecta limit to a common
distribution at high velocities because the blast wave forgets its
history as it approaches the stellar surface. We present formulae for
the final density distribution of this material as a function of mass,
for both radiative and efficiently convective envelopes.  These
formulae limit to the well-known planar, self-similar solutions for
mass shells approaching the stellar surface. However, the assumption
of adiabatic flow breaks down for shells of low optical depth, so this
planar limit need not be attained. The event of shock emergence, which
limits adiabatic flow, also produces a soft X-ray burst of
radiation. Formulae are given for the observable properties of this
burst and their dependence on the parameters of the explosion.
Motivated by the relativistic expansion recently inferred by Kulkarni
et al. (1998) for the synchrotron shell around SN1998bw, we estimate
the criterion for relativistic mass ejection and the rest mass of
relativistic ejecta.

We base our models for the entire ejecta distribution on the
high-velocity solution, on our shock velocity formula, and on
realistic radiation pressure distributions. We also present simpler,
but less flexible, analytical approximations for ejecta distributions.
We survey the ejecta of the polytropic hydrogen envelopes of red
supergiants.  Our models will be useful for studies of the light
curves and circumstellar or interstellar interactions of core-collapse
supernovae, and the birth of pulsar nebulae in their ejecta.
\end{abstract}

 
\section{Introduction} \label{S:intro} 

All stars of initial mass greater than about $8M_\odot$ end violently
when their cores collapse; many or all of these produce supernova
explosions. Supernovae are of paramount importance to the evolution of
galaxies: they form new elements and recycle the products of stellar
evolution. Their energy is converted into motion and heat,
affecting the dynamical state of the surrounding gas; indeed, they are
responsible for much of the hot, X-ray emitting gas in galaxies, and
for the formation of superbubbles and outflow in starburst
galaxies. Observationally, supernovae provide a natural laboratory in
which to test theories of stellar evolution and of the circumstellar
environments of massive stars: supernova 1987A is a notable example
(\cite{M93}).

How does the stellar structure of a progenitor affect the aftermath of
its explosion?  The observable properties of a core-collapse supernova
are determined by the distribution of material and temperature in its
freely-expanding ejecta. The supernova's optical display begins with a
photospheric stage, in which a recombination front moves inward
through the expanding ejecta. This process is pre-determined by the
temperature, density and opacity profiles of the outcast material. At
some point, the energy from radioactive decays begins to dominate the
light curve. The nebular display is determined by the distributions of
both decaying isotopes and gamma-ray opacity. Later, as the ejecta
collide with circumstellar or interstellar material, two shocks are
born at the interface: a forward shock accelerating the surrounding
material, and a reverse shock decelerating the ejecta (\cite{M74},
\cite{C82}, \cite{DC97}, \cite{TM98}). This interaction depends
critically on the distributions of ejected and ambient material. It is
responsible for the appearance of the supernova in the radio, X-ray
and infrared bands for centuries after the explosion.

The distributions that characterize a supernova's ejecta are, in turn,
determined uniquely by the structure of the progenitor star (although
the distribution of radioactive elements is an added complication).
The relation between progenitor and ejecta is established by the
hydrodynamics of the supernova explosion itself, which occurs in three
stages. In the first stage, a shock propagates through the stellar
material to be ejected, pressurizing it and setting it into motion: we
call this the \emph{blastwave} stage, following Chevalier (1976). Once
the shock arrives at the stellar surface, a \emph{rarefaction} phase
begins: the material accelerates as its heat is converted into kinetic
energy. This stage (the most difficult to model) ends when
acceleration ceases, once the ejecta's pressure is too slight to be
dynamically important. In the \emph{ejecta} stage, the outcast
material was coasts into homologous expansion, with a velocity
distribution ${\bf v} = {\bf r}/t$.

The correspondence between progenitor and ejecta can be established
numerically on a case-by-case basis. This approach has been pursued
for many years (e.g., \cite{FA77}). However, many aspects of the late
stages of stellar evolution are uncertain, including rotation and
pulsations (Heger et al. 1997), nuclear reaction rates, mass loss
prescriptions and models for semiconvection (\cite{WLW93}), and it is
likely that interactions with stellar (e.g., Nomoto, Iwamoto, \&
Suzuki 1995) or substellar companions will affect supernova
progenitors. Numerical simulations may describe specific cases quite
accurately, but only limited insight into the physical relationship
between progenitor and ejecta can be derived from an individual
case. 

Supernova progenitors and their ejecta display strikingly similar
features, when the initial and final densities are considered
functions of a Lagrangian variable, like the enclosed mass. 
Does this imply that there is a simple relationship between the
two? This paper will investigate the existence of such a relationship,
that is, the degree to which the variables that describe a mass shell
in the progenitor control its properties after the explosion
happens. A local relation between the initial and final variables
implies that the physics of the explosion can be understood in simple
terms, despite the complicated hydrodynamical interactions
involved. The pressure-based models of \S \ref{S:pressuremodels}
demonstrate that this is the case. 

Another goal of this paper will be to provide simple models for broad
classes of supernova progenitors, like red supergiants. When
progenitor stars are similar in structure, their ejecta will also take
similar distributions. A simple model that can reproduce these ejecta
accurately enough to show the differences between them, like the
harmonic-mean models of \S \ref{SS:harmRSGs}, should prove valuable
for studies of supernova light curves and circumstellar interactions.

\subsection{Previous investigations}
The analytical study of supernova hydrodynamics has relied on familiar
self-similar analogs. Over a considerable fraction of the progenitor,
the density distribution resembles a power law in radius:
$\rho_0\propto r_0^{-n}$ with $n\sim 1.5-3$ above the core, but
usually with a shallower distribution further out. In a strictly
power law, spherical distribution, the blastwave stage takes the
self-similar form given by Taylor (1950) and Sedov (1959). Chevalier
(1976) likened the structure of a blastwave in an increasingly steep
density distribution to a progression of Sedov blastwaves in different
power law media, and concluded that a detached shell (density maximum)
would occur if density became steeper than a critical value. By
analogy to the Sedov blastwave with $\rho_0\propto r_0^{-3}$, Bethe \&
Pizzochero (1990) conjectured that the shock in supernova 1987A would
have nearly a constant velocity. Chevalier \& Soker (1989)
approximated the central progenitor distribution as $\rho_0\propto
r_0^{-17/7}$, which produces a particularly simple blastwave
distribution (the Primakoff-Sedov form) in which density and pressure
are power laws in radius and time.

The resemblance to a Sedov blastwave must break down as the shock
approaches and emerges from the stellar surface, and the rarefaction
phase begins. But there is an analytical solution to guide the
understanding of this process as well. The density varies as a power
of depth in the outer layers of a star, and radial curvature can be
neglected near enough to the stellar surface.  A planar version of
this problem was initially considered by Gandel'man \&
Frank-Kamenetsky (1956) and solved by Sakurai (1960). In this context,
the shock front accelerates as it runs down the declining density
gradient, and material expands into vacuum after the shock emerges
from the surface. The ratio of final to postshock velocity is the same
for all fluid elements in this solution; the density and pressure
approach steep power laws in velocity as the material expands
freely. This solution has been applied to the pulse of emission
associated with shock breakout and the early phase of the light curve
(\cite{C92}), as well as the distribution of high-velocity ejecta from
the explosion (e.g., \cite{CS89}, \cite{IN89}, \cite{LN90}). Imshennik
\& Nad\"ezhin (1989) and Chevalier (1992) showed that spherical
curvature should complicate matters, because a mass shell in the
planar solution must travel a long way (compared to its initial depth)
to reach its final velocity. Kazhdan \& Murzina\footnote{We are
grateful to the referee, Roger Chevalier, for bringing this work to
our attention.} (1992) took the dynamical effect of sphericity into
account in a first-order perturbation analysis of Sakurai's
solution. In this paper, we will find the correction of the planar
solution to higher order by a different method.

Sakurai's planar solution describes the end of shock motion and the
beginning of rarefaction. But even with the first-order effect of
sphericity accounted for, this solution only holds for the
highest-velocity ejecta, or for the brief period in which the shock
crosses and ejects a thin outer shell. It does not describe how this
brief period joins onto the blastwave phase that preceded it and the
rarefaction phase that follows it, in which the bulk of the ejecta
approaches its final velocity. The rarefaction phase is very
complicated in general, even if the blastwave was one of the Sedov
forms; rarefaction is therefore a formidable problem.

There is one counter-example. For the special case of the Primakoff
blastwave, in which density and sound speed are power laws of radius
and time, a rarefaction wave will travel along a power law
trajectory. Chevalier \& Soker (1989), using the Primakoff form to
approximate the blastwave in SN1987A, argued that the spherical
rarefaction process will be self-similar in this case. Then, the ratio
between the final velocity and the velocity at the rarefaction wave
should be the same for any mass shell. This simple argument specifies
that the density should settle into a shallow power law of velocity,
in the inner region of the ejecta. Joining this shallow inner power
law with the steep outer power law specified by Sakurai's planar
solution, and imposing overall mass and energy conservation, Chevalier
\& Soker found the velocity of the break between the two
behaviors. Although it is based on idealizations of the flow, this
model compares admirably with simulations involving blue supergiant
progenitors.

\subsection{This work}

In this paper we address each stage of the expulsion of stellar
envelopes in core collapse supernovae, in order to arrive at a
reliable model for the final ejecta distribution that can accommodate
possible variations in progenitors' stellar structures. Wherever
possible, we make use of scaling relations to simplify the
problem. When considering red supergiants we concentrate primarily on
the outermost composition layer (the hydrogen envelope) where our
models will be most accurate, approximating these envelopes as
polytropes when it is useful to do so. Table \ref{t:models} gives an
overview of the models developed in this paper; these are motivated by
and then checked against simulations for which the initial
configuration is the result of a stellar evolution study. These
progenitors were kindly supplied to us by Stan Woosley and Ken'ichi
Nomoto.

We begin by examining in \S \ref{S:scaling} the scaling relations that
connect the hydrodynamical behavior of similar initial structures,
under the idealization of adiabatic flow. The scaling to similar
solutions will be especially useful in \S \ref{S:highv}, where we
present solutions for the high-velocity ejecta of convective and
radiative envelopes with a single form for each.

Section \ref{S:progenitors} discusses the general features of
supernova progenitor stellar structures. These are marked by
composition zones of vastly different density and radial scale. The
blastwave shock typically decelerates inside each layer, and
accelerates in the steep density declines between zones. The shock
wave accelerates as it approaches the surface of the star, at the end
of the blastwave phase. 

The models we develop will require an approximation for the detailed
motion of the blastwave shock front. The forward shock defines the
outer boundary of the flow in the blastwave stage. Moreover, the
blastwave shock velocity sets the entropy distribution for the ensuing
adiabatic flow (for material not subject to reverse shocks). The shock
velocity thereby affects the rarefaction phase as well, and the
relation between pressure and density in the final ejecta phase. The
shock velocity is also necessary to set the coefficients of the power
laws that describe the highest-velocity ejecta.  In reality, the shock
velocity must be determined self-consistently by matching it to the
postshock pressure in the blastwave phase. Previous studies
(Kompaneets 1960, Laumbach \& Probstein 1969, Whitham 1974, Klimishin
\& Gnatyk 1982, Koo \& McKee 1990) have suggested a number of formulae
for the velocities of shocks in inhomogeneous, three-dimensional
distributions; all of these formulae have drawbacks in the context of
supernovae, and we will suggest another simple shock velocity formula
better adapted to this context in \S \ref{S:vshock}. This shock
velocity approximation is the product of well-known scaling laws for
spherical decelerating shocks and planar accelerating shocks. Despite
its simplicity, this approximation tracks the successive periods of
shock acceleration and deceleration with remarkable accuracy. This
approximation gives the entropy distribution in the ejecta, except
where a reverse shock interferes. It also gives the coefficients that
describe the highest-velocity ejecta.

In \S \ref{S:highv} we address the dynamics of material near the
surface in the progenitor, the material that is ejected at high
velocity. The initial distribution of this material, described in \S
\ref{SS:rho01}, has a predictable form in radiative and polytropic
envelopes.  Having a reliable estimate of the velocity of the shock
that launches this material, we calculate in \S \ref{SS:ssresults} the
coefficients of the power laws that describe its ejecta, as given by
the planar, self-similar solution (Gandel'man \& Frank-Kamenetsky
1956, Sakurai 1960; see the Appendix). Most material attains a smaller
velocity than the planar formula would give, because spherical
expansion reduces the pressure below its planar value. This is a
geometrical effect, however, and because the initial density
distributions of this material are similar among broad classes of
envelopes (radiative or convective), the final ejecta distributions
are also similar. In \S \ref{SS:commonform}, we present simple
formulae that capture the deviation of important quantities away from
their self-similar power laws.

The self-similar solutions that give power laws for the high-velocity
ejecta are not valid to infinite velocity, of course. The assumption
of adiabatic flow breaks down for material involved in shock breakout,
because radiation diffuses or streams out of the gas. This process
also produces a burst of soft X-rays. In \S \ref{SSS:speedlimit} we
present formulae for the upper velocity limit for adiabatic flow. We
use our non-relativistic analysis to derive an approximate criterion
for relativistic mass ejection and the rest mass of relativistic
ejecta (if any) in \S \ref{SSS:relativity}. In \S \ref{SSS:emergence},
we present formulae for the characteristics of the X-ray
burst. Although previous authors have treated the radiation outburst
in a similar manner (e.g., \cite{IN89}), the shock velocity formula of
\S \ref{S:vshock} allows us to elucidate the effects of stellar
structure on the outburst.

The results of \S \ref{S:highv} apply to the high-velocity ejecta,
those at velocities a few times greater than the systemic
velocity. However, little of the mass or energy in the ejecta lies in
this velocity range. For the bulk of ejecta, the details of the
progenitor's structure break the scaling relations that allow the
high-velocity ejecta to be given accurately by simple formulae. We
note that the final \emph{pressure} distribution is always smooth and
continuous in the ejecta, even if the density distribution is not. For
this reason, we base our models for the entire ejecta distribution,
discussed in \S \ref{S:pressuremodels}, on models for the pressure
distribution that mimic the smooth pressure functions seen in
simulations. We find a simple way to extend the pressure distributions
derived in \S \ref{SS:commonform} inward in mass.  The final density
distribution is specified by the combination of this pressure model
with the entropy left behind in the shock. These pressure-based models
constitute a \emph{local} theory, in that they relate the final
density of a mass shell with its initial state. These models fail in
regions that are traversed by a strong reverse shock, because we do
not account for the entropy it deposits; this failure is worst in the
inner composition layers of red supergiants. Even so, there is a jump
in density between the ejected helium mantle and the ejected hydrogen
envelope that is not affected by the reverse shock; a simple analysis
of this phenomenon is presented in \S \ref{SS:rhojump}.

Section \ref{SS:harmRSGs} presents a different type of model for ejecta
distributions, in which the shape of the final density is described as
a simple, harmonic-mean interpolation between two power laws of
velocity. This is a \emph{global} model, since it gives an ejecta
distribution (in terms of three free parameters) without referring to
the detailed structure of the progenitor. For blue supergiants, we
adopt the theory of Chevalier \& Soker (1989) to motivate the
choice of parameters. For red supergiants, we find that the
harmonic-mean model can fit the ejected hydrogen quite accurately, but
it does not fit the ejected mantle because of the density jump between
the two regions.

In \S \ref{S:RSGs} we develop simplified polytropic models for red
supergiant progenitors of type II supernovae. We argue that these can
be adequately described by two parameters that are ratios of masses in
the progenitor. Moreover, these mass ratios are tightly correlated,
so that there is essentially a one-parameter family of red supergiant
progenitors. We investigate this family of models numerically to find
the distribution of the free parameter used in the pressure-based
model of \S \ref{S:pressuremodels} and the three free parameters in 
the harmonic-mean models of \S \ref{SS:harmRSGs}; results are given in
Table \ref{t:harmonics}. 

Section \ref{S:simcomp} contains a comparison between our models and
the ejecta of simulated supernovae whose progenitors are the end
products of stellar evolution calculations. This comparison allows a
tabulation of the errors introduced in the pressure-based and
harmonic-mean models (Table \ref{t:errors}) in a realistic context. 
The models developed in this paper help to explain many of the
features seen in numerical simulations of supernovae. This is true
even for features are not reproduced accurately by the models, those
caused by reverse shocks or by variations in the polytropic index at
the outside of the progenitor. These features are discussed in \S
\ref{S:simcomp} and in the Conclusions. 

In the Appendix, we develop a fully Lagrangian treatment of the
self-similar shock acceleration problem solved by Sakurai. This
problem has previously been addressed in two steps, with an Eulerian
solution prior to emergence matched to a Lagrangian description of the
subsequent expansion. The fully Lagrangian description facilitates
analyses of the highest-velocity material. We use it to compare our
results of \S \ref{SS:commonform} to the suggestion of Litvinova \&
Nad\"{e}zhin (1990) that spherical expansion essentially truncates
planar, self-similar acceleration, and we find that this happens at
about three stellar radii.

\section{Scaling to similar solutions} \label{S:scaling}

Although the phenomenon of core collapse is quite complex (e.g.,
\cite{C95}), the resulting blastwave in the stellar envelope has
relatively simple dynamics. Gravity is relatively insignificant except
for the interior layers, which may suffer infall. The gain or loss of
energy to nuclear reactions is also dynamically insignificant except
for these inner layers, although they are probably mixed to
higher velocities by the Rayleigh-Taylor instability. (Heat from
decaying ${\rm {}^{56} Ni}$ may blow bubbles in the ejecta [Arnett,
Fryxell \& M\"uller 1989; Li, McCray \& Sunyaev 1993]). Likewise,
neutrino heating is negligible outside the dense circumnuclear
material. The pressure is radiation dominated, and the ratio of
radiation to gas pressure remains constant (for each mass shell)
throughout the period of adiabatic motion. The radiation diffuses
sufficiently little during the period of acceleration that the flow
can be modeled as adiabatic, although this approximation fails for the
optically thin shell associated with the UV or X-ray flash (e.g.,
Ensman \& Burrows 1992). The heat lost to radiation never exceeds a
few percent of the total explosion energy, implying that radiative
transfer can be neglected in general.

Under the approximation that the flow is adiabatic,
radiation-dominated and non-gravitating, its only dimensional scales
are the explosion energy ($E_{\rm in}$), \mbox{\emph{ejected}} mass ($M_{\rm ej}$)
and initial stellar radius ($R_\star$). Explosions with different
values for these parameters, but the same initial density profile, are
identical after rescaling the hydrodynamic variables by their
characteristic values. Common stellar density profiles recur on a
variety of scales, and the resulting (normalized) ejecta distributions
are identical.  We will make use of this principle in \S
\ref{S:highv}.

We take a mass coordinate $m(r)\equiv M(r)-M_{\rm rem}$ (where $M_{\rm
rem}$ is the mass of the material that will become the remnant) that
is zero at the boundary between ejected and remnant material, and
increases to $M_{\rm ej}$ at the exterior. The configuration of the
progenitor is specified by the initial density $\rho_0(m)$ and the
initial radius $r_0(m)$. If we normalize to the characteristic scales
of mass, radius, velocity, time, density, pressure and entropy, the
flow expressed in normalized variables evolves independently of these
scales.  The basic characteristic scales are $M_{\rm ej}$, $R_\star$ and
$E_{\rm in}$, from which it is useful to derive the following characteristic
quantities:
\begin{eqnarray} \label{eq:scales}
v_\star &\equiv& \sqrt{E_{\rm in}/M_{\rm ej}} = 2240
	 \left(\frac{E_{\rm in}}{10^{51} {\rm erg}}\right)^{1/2}
	 \left(\frac{M_{\rm ej}}{10 M_\odot}\right)^{-1/2}
	{\rm km\, s^{-1}}, \nonumber \\ 
t_\star &\equiv& R_\star \sqrt{M_{\rm ej}/E_{\rm in}} = 1.552 \times 10^4
	\left(\frac{R_\star}{50 R_\odot}\right) 
	 \left(\frac{E_{\rm in}}{10^{51} {\rm erg}}\right)^{-1/2}
	 \left(\frac{M_{\rm ej}}{10 M_\odot}\right)^{1/2} {\rm s}, \nonumber\\
\rho_\star &\equiv& \frac{M_{\rm ej}}{R_\star^3} =  4.720 \times 10^{-4}
	 \left(\frac{M_{\rm ej}}{10 M_\odot}\right)
	\left(\frac{R_\star}{50 R_\odot}\right)^{-3} {\rm g\,cm^{-3}}, \nonumber\\
p_\star &\equiv& \frac{E_{\rm in}}{R_\star^3} = 2.373 \times 10^{13}
	 \left(\frac{E_{\rm in}}{10^{51} {\rm erg}}\right)
	\left(\frac{R_\star}{50 R_\odot}\right)^{-3} {\rm erg\,cm^{-3}}, \nonumber\\ 
s_\star&\equiv&\frac{E_{\rm in}^{3/4}R_\star^{3/4}}{M_{\rm ej}} = 2.278\times 10^{13}
	 \left(\frac{E_{\rm in}}{10^{51} {\rm erg}}\right)^{3/4}
	\left(\frac{R_\star}{50 R_\odot}\right)^{3/4}
	 \left(\frac{M_{\rm ej}}{10 M_\odot}\right)^{-1} {\rm
	 erg^{3/4}\,cm^{3/4}\,g^{-1}}.\nonumber\\ 
\end{eqnarray}
We will call the quantity $s\equiv p^{1/\gamma}/\rho$ the
`entropy'; for $\gamma=4/3$, it is proportional to the phase space
volume and the photon-baryon ratio. 

The goal of this paper is to provide formulae for the end product of
an adiabatic explosion. In the final state of adiabatic free
expansion, the velocity of a given mass shell is fixed, while its
density and pressure fall off as $t^{-3}$ and $t^{-3\gamma}=t^{-4}$,
respectively. The ejecta properties scale with the ejected mass,
stellar radius and blastwave energy: for a given feature in the ejecta
(the half-mass shell, say), 
\begin{eqnarray}\label{eq:scalings}
v_f&\propto& v_\star=E_{\rm in}^{1/2}M_{\rm ej}^{-1/2}
\nonumber \\ 
 \rho_f&\propto& \rho_\star(t/t_\star)^{-3}=E_{\rm in}^{-3/2}
M_{\rm ej}^{5/2} t^{-3}
\nonumber \\ 
p_f&\propto& p_\star (t/t_\star)^{-4} =E_{\rm in}^{-1} 
M_{\rm ej}^2 R_\star t^{-4}
\end{eqnarray}
These overall scalings are very important for inferring gross
quantities, but they say nothing about the relation between the
progenitor's structure and the distribution of its ejecta.

So, we are interested here in the normalized ratios $v_f/v_\star$,
$\rho_f t^3/\rho_\star t_\star^3$, and $p_f t^4/p_\star t_\star^4$ as
parameterized, for instance, by the normalized mass $m/M_{\rm ej}$. Note
that these are `final' values only in the sense of an adiabatic
explosion into vacuum: the radiation pressure will differ from $p_f
t^4$ because of radiation diffusion and radioactive decay, and the
distribution of matter in the ejecta ($v_f(m)$ and $\rho_f(m) t^3$)
will ultimately be destroyed as interstellar material is swept up.

Throughout this paper, we use the normalized mass coordinate $\m\equiv
m/M_{\rm ej}$, which varies from 0 to 1 in the ejecta. Because all of the
hydrodynamical variables approach power laws in the limit of high
velocity, $\m\rightarrow 1$, we introduce a notation that removes this
dependence. In this limit, a variable $x$ varies as
$x\propto(1-\m)^{w_x}$ for some exponent $w_x$; so, we define
$f_x(\m)$ by,
\begin{equation}\label{eq:fnotation}
x = x_\star f_x(\m) (1-\m)^{w_x}, 
\end{equation}
where $x_\star$ is a characteristic value for $x$ (a combination of
$E_{\rm in}$, $M_{\rm ej}$ and $R_\star$). The exponents $w_x$ and limiting
coefficients $f_x(1)$ can be evaluated by combining the initial
density and shock velocity in the subsurface layers with results from
the planar, self-similar solution for the ejection of this material;
this is done in \S \ref{SS:ssresults}. Unless otherwise noted, $f_x$
refers to the final value of $x$ in free expansion. So, $f_v$ refers
to $v_f$, whereas $f_{v_s}$ refers to $v_s$, the shock velocity.

\section{Stellar structures of the progenitors: stellar models and
approximations} \label{S:progenitors} 

The progenitors of core collapse supernovae are stars initially more
massive than about $8M_\odot$, whose cores have advanced though carbon
burning. The degenerate iron core approaches the Chandrasekhar mass in
the subsequent burning stages, the process of `core convergence'
(Arnett 1997). This core has the radius and density of an iron white
dwarf. The circumnuclear shells decrease in density away from this
core, characteristically as $\rho_0\propto r_0^{-3}$. Outside the carbon
shell, the helium envelope begins. There is typically a steep,
shelf-like decline in the density between the carbon and helium. The
helium envelope is typically radiative prior to the explosion
(\cite{A97}).

Above the helium envelope sits whatever hydrogen-rich envelope has
survived the stellar mass loss, which may be none at all. The
structure of the star interior to the hydrogen is little altered by
its presence or absence, because the hydrogen burning shell has a
pressure and temperature that is negligible relative to the interior
(\cite{A97}).  In stars that have hydrogen, the boundary between
hydrogen and helium is marked by a density drop even steeper and
greater than at the base of the helium envelope. Presupernova stars
with radiative hydrogen envelopes are blue supergiants (BSGs); those
with convective hydrogen envelopes are red supergiants (RSGs); and
those with no hydrogen at all are helium or carbon
stars. 

Core-collapse supernovae are classified observationally into types II,
Ib, and Ic. Type II supernovae show hydrogen lines, type Ib show
helium but no hydrogen, and type Ic show neither. Because the spectrum
may change with time, supernovae can shift in classification
(Filippenko 1997): supernova 1993J was one such (e.g., Shigeyama et
al. 1994). Thus red and blue supergiants have the capacity to produce
types II and Ib, helium stars can produce types Ib and Ic, and carbon
stars can only produce type Ic.

We will concentrate on the topmost stellar envelope, which we will
term the \emph{outer envelope} and whose mass we designate $M_{\rm
env}$. This layer is the hydrogen-rich envelope of those stars that
have hydrogen. Those layers that lie interior to the outer envelope,
but are also ejected in the explosion, we will refer to as the
\emph{mantle}, with mass $M_{\rm man}$. The total ejecta mass $M_{\rm
ej}$ is the sum of these, $M_{\rm ej}=M_{\rm man}+M_{\rm env}$. The
\emph{remnant}, with mass $M_{\rm rem}$, is the stellar material that
is not ejected; this is slightly greater than the mass of the relict
object because of rest energy radiated in the explosion. The total
progenitor mass is $M_\star= M_{\rm rem} + M_{\rm ej}$. Our focus on
the outer envelope is motivated by the observation that it receives
the lion's share of the explosion energy; moreover, our simulations
indicate that the distribution of the outer envelope ejecta is
determined primarily by its own structure and the mass $M_{\rm man}$
of the ejecta it overlies, not appreciably by the structure of the
underlying mantle material.

Convective envelopes, like those in RSGs, can be approximated as
$n=3/2$ polytropes in structure. This approximation would be exact if
the convection were efficient, but the presence of a superadiabatic
gradient breaks this idealization at large initial radii (Nomoto \&
Sugimoto 1972). Models of convection can prescribe a superadiabatic
gradient, but the results depend on the model used (mixing length, for
instance), as well as stellar parameters (such as $L_\star/L_{\rm
edd}$). We will idealize convective outer envelopes as $n=3/2$
polytropes, but the reader is advised to keep this caveat in mind.

Radiative envelopes of constant (Thomson-dominated) opacity, such as
those in BSGs, limit to an effective polytropic index of $n=3$ toward
the outside of the star, because such envelopes behave like $n=3$
polytropes when luminosity and mass are proportional. However, the
interior of blue supergiants is better described by a lower effective
polytropic index. In the BSG progenitor of Nomoto, we find that
$2.1<n_{\rm eff}<2.4$ for the inner $75\%$ of the ejected mass, with
$n_{\rm eff}$ climbing to $3$ in the exterior $25\%$. (A power law of
index $17/7=2.43$ was adopted by \cite{CS89}.)

The structure of a polytropic envelope is determined, up to overall
scaling factors, by a single parameter, corresponding to the fact that
the Lane-Emden equation reduces to a first-order differential equation
in terms of homology-independent variables (\cite{C39}). Which
structure is appropriate is determined by the inner boundary
condition, defined by the mantle's mass ratio $(M_{\rm man}+M_{\rm
rem})/M_\star$ and the radius ratio $R_{\rm man}/R_\star$. In order to
get a single parameter from these two, consider that the Lane-Emden
equation describing the outer envelope can be integrated all the way
to the center of the star. This hypothetical integration through the
region actually occupied by the mantle will yield a polytrope that
behaves as if it lay above a point mass at the origin. The value of
the point mass compared to $M_\star$, which we term $q$, provides a
single parameter with which to distinguish outer envelope structures.

To understand the relation between $q$ and the more familiar
parameters $(M_{\rm man}+M_{\rm rem})/M_\star$ and $R_{\rm
man}/R_\star$, we must examine the central behavior of
polytropes. Polytropes with $n<3$ limit as $r\rightarrow 0$ to the
central distribution (\cite{C39}),
\begin{equation}
\rho_0\propto (BR_\star/r_0-1)^n,
\end{equation} 
for some constant $B\geq 1$ (equality when $q=1$). In red supergiants
($n=3/2$), one can typically ignore the envelope mass taken up by the
volume of the mantle, both because the mantle radius is very small
($R_{\rm man}\simeq 1\% R_\star$), and because the envelope mass is
not highly concentrated towards the center ($\m \propto r_0^{3/2}$). So
for RSGs, we can use the core-envelope mass ratio to parameterize the
envelope structure:
\begin{equation}
q \simeq 1-M_{\rm env}/M_\star. 
\end{equation} 
In BSGs, an effective value of $q$ is less easily determined, because
of the variation in the effective polytropic index $n_{\rm eff}$. 

Polytropes with $q=0$ extend continuously and smoothly to the origin,
where $d\rho/d r=0$ (because $dp/dr=0$ at the center). For any value
of $q$ greater than zero, the central density profile is $\rho \propto
r^{-n}$, as long as $n<3$ (\cite{C39}). In polytropic envelopes that
are much more massive than their cores, so that $q\ll 1$, this
behavior is limited to a small region around the center (of similar
mass), outside of which the structure returns to the flatter, $q=0$
form. On the other hand, envelopes with $q\lesssim 1$ are dominated by
the gravity of their cores: the density profile is everywhere at least
as steep as the central behavior. 

\subsection{Outer density distributions of stars}\label{SS:rho01}

We will be especially interested in the region of a progenitor nearest
the surface, because it is in this layer that the shock accelerates
and casts away the highest-velocity ejecta. Fortunately, the equations
of polytropic stellar structure are simplified for this material. 
The hydrostatic equation for a polytropic envelope can be integrated
to give: 
\begin{equation}\label{eq:rhoform}
\rho_0^{1/n} = \rho_1^{1/n} \left(\frac{R_\star}{r_0}-1\right)
	\frac{\int_{R_\star^{-1}}^{r_0^{-1}} [m(r)+M_{\rm rem}] dr^{-1}}
	{\int_{R_\star^{-1}}^{r_0^{-1}} M_\star dr^{-1}}, 
\end{equation} 
where 
\begin{equation}\label{eq:rho1defn}
\rho_1^{1/n} \equiv \frac{GM_\star}{(n+1)KR_\star},
\end{equation}
and $P_0 = K\rho_0^n$.  In a shallow (outer) layer whose mass is
negligible compared to $M_\star$, the ratio of mass integrals is
approximately unity. Therefore, this region has an analytical density
structure:
\begin{equation}\label{eq:outerform}
\rho_0 = \rho_1 \left(\frac{R_\star}{r_0}-1\right)^n. 
\end{equation} 
This is valid only for masses such that the enclosed mass can be
approximated by $M_\star$; however, the region of validity may be
relatively large in radius. For light envelopes whose entire mass is
negligible ($q=1$), this distribution holds for the entire envelope.
Here, $n$ is $3/2$ for efficiently convective envelopes (although
inefficient convection may invalidate this form) and exactly $3$ for
radiative envelopes of constant opacity (an analytical result in the
outermost layers, where the ratio $L/M$ is constant).

The stellar density coefficient $\rho_1$ is related to the mass
coefficient $f_{\rho_0}(1)$, defined in equation (\ref{eq:fnotation}),
by the relation
\begin{equation}\label{eq:rho1f1}
f_{\rho_0}(1)^{n+1} = \frac{\rho_1}{\rho_\star}
 \left(\frac{n+1}{4\pi}\right)^n; 
\end{equation}
the corresponding exponent for $\rho_0$ as a power law in $(1-\m)$
is, 
\begin{equation} 
w_{\rho_0} = \frac{n}{n+1}. 
\end{equation} 

For radiative envelopes of constant opacity, 
\begin{equation}\label{eq:radrho1}
\rho_1 = \frac{a (\mu m_H)^4}{192 k_B^4} \frac{G^3 M_\star^3}{R_\star^3}
\frac{\beta^4}{1-\beta}, 
\end{equation} 
where $1-\beta \equiv L_\star/L_{\rm Edd}$ is the ratio of the stellar
luminosity to the Eddington limit, or equivalently, $1-\beta$ is the
ratio of radiation pressure to total pressure in the outermost layers
of the star.  From equation (\ref{eq:rho1f1}), for radiative stars,
\begin{equation}
f_{\rho_0}(1) = 0.148 \left(\frac{M_\star}{M_{\rm ej}}\right)^{1/4}
\frac{1-\beta}{\beta^{1/4}}
\left( \frac{\mu}{0.62} \right)
\left( \frac{M_\star}{10 M_\odot} \right)^{1/2}, 
\end{equation}
and for Thomson opacity, 
\begin{equation} 
1-\beta = \left(\frac{L_\star}{3.86 \times 10^5 L_\odot}\right)
\left(\frac{M_\star}{10 M_\odot}\right)^{-1}
\left(\frac{1+X_H}{1.7}\right). 
\end{equation} 
For convective envelopes with $0<q<0.9$ we find that,
\begin{equation}\label{eq:frho01conv}
f_{\rho_0}(1) \simeq \left(\frac{M_{\rm env}}{M_{\rm ej}}\right)^{2/5}
\left(0.37-0.18q+0.096q^2\right). 
\end{equation}
This fit to the Lane-Emden solution is accurate to within 1\% for the
given range of $q$.

In summary, each layer of a star has a density distribution that is
$\rho_0 \propto r_0^{-n}$ at its interior, and that may flatten out
towards constant density before dropping precipitously near the edge
of the shell. The appropriate value of $n$ is $3/2$ for efficiently
convective envelopes and $2.1-3$ for radiative envelopes. Besides $n$,
the structure of a polytropic envelope depends on a central
concentration parameter $q$; for the outer envelopes of RSGs, $q$ is
essentially the mass interior to the envelope, divided by the
progenitor mass.  There are sharp drops in density between the carbon
and helium and between the helium and hydrogen shells; the latter is
most drastic in the case of red supergiants.

\section{Model for shock propagation} \label{S:vshock} 

It is through this layered sequence of envelopes that the supernova
shock propagates. A model for the shock velocity must account for the
general deceleration as mass is swept up, punctuated by accelerations
as the shock encounters steep density gradients. The velocity of the
blastwave shock is clearly affected by all of the material it has
encountered, as well as the conditions at the shock front. But, the
history of analytical shock velocity approximations has primarily been
the search for a combination of \emph{local} variables that controls
(or at least mimics) the motion of the front. Formulae in the
literature have typically given $v_s$ in terms of the local initial
density $\rho_0$ and radius $r_0$, as well as the blastwave energy,
$E_{\rm in}$ (an integral quantity).  

\subsection{Previous models} \label{SS:prevvshock}

Kompaneets (1960) considered shock acceleration in the case of an
initially exponential density distribution. He approximated the
postshock pressure as a constant fraction of the mean energy density,
leading in the case of spherical symmetry and in our notation to the
expression,
\begin{equation}\label{eq:vsK60}
\frac{v_s}{v_\star} = \Gamma_{\rm K60} 
\left(\frac{\rho_0}{\rho_\star}\right)^{-1/2}
\left(\frac{r_0}{R_\star}\right)^{-3/2}, 
\end{equation}
where $\Gamma_{\rm K60}$ is a constant that can be evaluated in
specific cases.  As discussed by Koo \& McKee (1990), this
approximation overestimates the shock velocity as the shock
accelerates, both because the mean pressure drops in relation to the
mean energy density, and because the postshock pressure drops in
relation to the mean pressure. Laumbach \& Probstein (1969) developed
an approximation for the motion of the shock wave in which the shocked
material is assumed to be concentrated in a thin shell behind the
shock, moving at the postshock velocity. The approximation involved
Taylor expansions for the form of this shell, and related these to the
motion of the shock. As Koo \& McKee (1990) discuss, this
approximation tends to underestimate the shock velocity, and fails for
strongly accelerating shocks.

Klimishin \& Gnatyk (1982) considered shocks that alternate between
periods of deceleration and acceleration, as shocks in supernovae
do. They proposed that the shock velocity should obey $v_s \propto
E_{\rm in}^{1/2} (\rho_0 r_0^3)^{-1/2}$ during periods of
deceleration, and $v_s \propto E_{\rm in}^{1/2} (\rho_0
r_0^3)^{-\beta_1}$ (with $\beta_1=0.2$) during periods of
acceleration. 

Koo \& McKee (1990), considering density distributions that steepen
monotonically, proposed a shock velocity that accelerates beyond a
characteristic radius. Specifically, they chose a postshock pressure
that changes from the form appropriate to spherical, decelerating
blastwaves (as in Sedov 1959) and that appropriate to strongly
accelerating blastwaves (as in Whitham 1974) at a transition radius
$R_{\rm s,t}$. This leads to the shock velocity formula,
\begin{equation}\label{eq:vsKM90}
\frac{v_s}{v_\star} = \Gamma_{\rm KM90} 
\left(\frac{\rho_0}{\rho_\star}\right)^{-\beta_1}
\left(\frac{r_0}{r_0+ R_{\rm s,t}}\right)^{-3/2}
\left(\frac{r_0 + R_{\rm s,t}}{ R_{\rm s,t}}\right)^{-\beta_2},
\end{equation} 
where $\Gamma_{\rm KM90}$ is a coefficient chosen to match Sedov-Taylor
blastwave if the central density distribution is flat. This
prescription worked well for the relatively smooth distributions
considered, which were flat at the origin and fell to zero density
asymptotically at large radii with a well-defined scale length.

However, supernova progenitors have layers of different compositions,
separated by shelf-like drops in density at the burning shells
(\S \ref{S:progenitors}). Therefore one must employ a model for shock
acceleration that does not rely on the choice of a single radial
scale. The ability to match the acceleration of the shock in the
interior shell boundaries and at the outer edge of the star is
particularly essential. 

\subsection{Improved model for shock velocity} \label{SS:vsmodel}

During a period of self-similar shock propagation (e.g., Sedov 1959),
the shock velocity obeys $v_s \propto \sqrt{E_{\rm in}/m(r_0)}$. During this
period, as well, the ratio $\rho(r_0)/\overline{\rho}(r_0)$ is a
constant, where $\overline{\rho}(r_0)$ is the average density of the
material to be ejected within $r_0$ (note that $\overline{\rho}$ and
$\m$ do not include the stellar core mass that will become the remnant
object). During a period of strong planar acceleration, when $r_0$ and
$m(r_0)$ are essentially constant (and so $\overline{\rho}(r_0)$ is
too), the shock velocity takes the form $v_s \propto [\rho(r_0)/
\overline{\rho}(r_0)]^{-\beta_1}$. We propose a continuous and simple
form.  To accommodate both spherical deceleration and planar
acceleration, we take the product form $v_s \propto \sqrt{E_{\rm in}/m(r_0)}
[\rho_0(r_0)/\overline{\rho}_0(r_0)]^{-\beta_1}$:
\begin{equation}\label{eq:vshock} \frac{v_s}{v_\star} = \Gamma
\m^{\beta_1-1/2} \left(\frac{\rho_0}{\rho_\star}\right)^{-\beta_1}
\left(\frac{r_0}{R_\star}\right)^{-3\beta_1}.
\end{equation} 
Using the approximation that a strongly accelerating shock runs along
a forward characteristic of the postshock flow, Whitham obtained
$\beta_1 \simeq 1/[2+\sqrt{2\gamma/(\gamma-1)}]$, or $0.2071$ for
$\gamma=4/3$. The self-similar, planar solution of Gandel'man \&
Frank-Kamenetsky (1956) and Sakurai (1960) gives $\beta_1=0.1909$ and
$\beta_1=0.1858$ for $n=3/2$ and $n=3$, respectively, so
$\beta_1\simeq 0.19$ is a reasonable approximation. (Note that in
writing equations [\ref{eq:vsK60}]--[\ref{eq:vshock}] for
$v_s/v_\star$ we have suppressed the explicit dependence $v_s\propto
E_{\rm in}^{1/2}$ that is important for blastwaves whose energy is not
conserved.)

It is possible to define $\Gamma(r_0)$ in terms
of moments of the distribution $\rho_0(r < r_0)$ in such a way as to
make eq. (\ref{eq:vshock}) strictly correct for self-similar
blastwaves. For a self-similar blastwave in the distribution
$\rho_0\propto r_0^{-k_\rho}$, equation (\ref{eq:vshock}) is correct if
$\Gamma = \sigma^{-1/2}[4\pi/(3-k_\rho)]^{-\beta_1}$, where $\sigma =
E_{\rm in}/(m v_s^2)$. For the Primakoff blastwave, $\sigma =
8/[3(\gamma+1)^2]$, so that,
\begin{equation}\label{eq:Gamma}
\Gamma=\frac{7}{\sqrt{24}}(7\pi)^{-\beta_1}
\end{equation} 
for $\gamma=4/3$.  Radiative stars roughly resemble the Primakoff
power law $\rho_0\propto r_0^{-17/7}$ (Chevalier \& Soker 1989), as do
the mantles of RSGs.  Despite the actual differences between stellar
distributions and this power law form, we find that together, equations
(\ref{eq:vshock}) and (\ref{eq:Gamma}) are remarkably accurate (Figure
\ref{fig:vshock}). 

Throughout the bulk of the stellar mantle and envelope in our
simulations, the exact choice of $\beta_1$ between the self-similar or
approximate values makes very little difference. However, the exact
values are important for determining the power laws of high-velocity
ejecta (\S \ref{SS:ssresults}). We find typical errors of $v_s$ in the
stellar envelope to be $\sim 2\%$, although higher ($\sim 10\%$) in
the mantle and in regions of strong acceleration. But, these errors
are in comparison to simulations of adiabatic, non-gravitating point
explosions in a gas sphere of mass $m(R_\star)=M_\star-M_{\rm
rem}$. In real supernovae, the shock velocity will differ
significantly from this formula at the base of the ejecta, due in part
to the mass that ultimately enters the remnant; in particular, the
shock velocity does not approach infinity as $m\rightarrow 0$,
contrary to equation (\ref{eq:vshock}). Our focus is on the envelope
and outer mantle, however, and for these equation (\ref{eq:vshock})
should remain accurate. 

Taking the value $\beta_1\simeq 0.19$ to represent both radiative and
convective envelopes, we can write equation (\ref{eq:vshock}) as:
\begin{eqnarray}\label{eq:AppxVshock}
v_s &\simeq& 0.794 v_\star \tilde{m}^{-0.31}
\left(\frac{\rho_0}{\rho_\star}\right)^{-0.19}
\left(\frac{r_0}{R_\star}\right)^{-0.57}\nonumber\\ 
&=& 0.794 \left(\frac{E_{\rm in}}{m}\right)^{1/2}
\left(\frac{m}{\rho_0 r_0^3}\right)^{0.19},
\end{eqnarray}
where $m$ is the enclosed ejecta mass.

To summarize the shock velocity approximations: the Kompaneets
approximation has $v_s\propto E^{1/2}\rho_0^{-1/2}r_0^{-3/2}$,
corresponding to a constant ratio of postshock to mean pressure. The
Koo \& McKee (1990) approximation goes from $v_s\propto
E_{\rm in}^{1/2}\rho_0^{-\beta_1}r_0^{-3/2}$ to $v_s\propto
E_{\rm in}^{1/2}\rho_0^{-\beta_1}r_0^{-\beta_2}$ at a transition radius
$R_{s,t}$. Klimishin \& Gnatyk took $v_s\propto E_{\rm in}^{1/2}(\rho_0
r_0^3)^{-1/2}$ and $v_s\propto E_{\rm in}^{1/2} (\rho_0 r_0^3)^{-\beta_1}$
in periods of deceleration and acceleration, respectively.  Our
approximation has $v_s\propto E^{1/2} m^{\beta_1-1/2
}\rho_0^{-\beta_1} r_0^{-3\beta_1}$. By	introducing $m$ and using
$\rho_0$ and $r_0$ only in the combination $\rho_0 r_0^3$, we avoid
having to choose dimensional scales. We compare the Kompaneets,
Klimishin \& Gnatyk and Koo \& McKee approximations to our equation
(\ref{eq:vshock}) in Figure (\ref{fig:vshock}).

Equation (\ref{eq:vshock}) represents an improved approximation for
$v_s$ in terms of the initial quantities $m$, $r_0$ and $\rho_0$ and
$E_{\rm in}$. These Lagrangian variables are especially convenient
because they are readily available in the initial state. 
Including $m$ allows the dimensionless ratio $\rho_0(r_0)/ \overline{
\rho}_0(r_0)$ to enter into the shock velocity approximation. Whereas the
shock velocity must in reality depend on the detailed structure of the
material encountered so far, it appears as if this single moment of
the initial density distribution is sufficient to specify the shock
velocity quite accurately. 

It is important to note that equation (\ref{eq:vshock}) is a purely
local theory for the shock velocity, in the sense that it only makes
reference to the variables $(m,r_0,\rho_0)$ at the same mass
shell in the progenitor. 

In the upcoming sections, we will be interested in relating the ejecta
pressure to its density by means of the `entropy', $s(\m) =
p^{1/\gamma}/\rho$. Using eq. (\ref{eq:vshock}) with $\gamma=4/3$, we
can specify the entropy of material that experiences no reverse shock:
\begin{equation} \label{eq:sshock} 
\frac{s(\m)}{s_\star} = \frac{6^{3/4}}{7^{7/4}} \Gamma^{3/2}
\left[\frac{\rho_0(\m)}{\rho_\star}\right]^{-(1+6\beta_1)/4}
\left[\frac{r_0(\m)}{R_\star}\right]^{-9\beta_1/2} 
\m^{(6\beta_1-3)/4}. 
\end{equation} 
This entropy formula, like the shock velocity formula from which it is
derived, is a local theory, invoking only $(m, r_0, \rho_0)$ for the
same mass shell of the progenitor. It falls short for shells
that will encounter a strong reverse shock, a weakness that will
prevent our models from describing the mantle ejecta
accurately. Including reverse shocks would require more information
than the local variables $(m,r_0,\rho_0)$ provide.

\section{High-velocity ejecta}\label{S:highv}
The analytical outer density distribution described in \S
\ref{SS:rho01} provides a stable platform from which to launch a study
of envelope expulsion in supernovae. The fact that polytropic
envelopes of the same index share this common form is useful for
understanding how a planar, self-similar analog is applicable to the
shock and rarefaction flow in this region. Moreover, this distribution
allows a characterization of the way in which the real, spherical flow
differs from the planar analog.

For material that is close enough to the stellar surface,
($R_\star-r_0\ll R_\star$), the density distribution of equation
(\ref{eq:outerform}) takes the limit
$\rho_0\rightarrow\rho_1(1-r_0/R_\star)^n$. If a mass shell stays in
this small range of radii throughout its period of acceleration,
spherical curvature will not affect its dynamics. Then, its velocity
will be set by the planar, self-similar problem first posed by
Gandel'man \& Frank-Kamenetsky (1956) and solved by Sakurai (1960),
which we reconsider in the Appendix. The material covered by this
solution settles into power law distributions whose indices are easily
derived once the shock velocity index $\beta_1$ is known
(c.f. \cite{CS89}). If the coefficients of density
[$\rho_1$ or $f_{\rho_0}(1)$] and shock velocity are known, the
coefficients of the final density distributions can also be derived
(c.f. \cite{IN89}). In \S \ref{SS:ssresults} the indices and
coefficients of this high-velocity material are calculated, using the
results of \S \ref{S:vshock} to set the shock velocity coefficient. 

This planar, self-similar analog is limited in its applicability to
real supernovae by several effects. For mass shells that differ
appreciably in radius from $R_\star$, $\rho_0$ is no longer a simple
power law in depth. For such shells, changes in radius are also sure
to make a planar approximation poor. In fact, as first mentioned by
Litvinova \& Nad\"ezhin (1990), curvature is actually negligible only
when $(1-r_0/R_\star)^{1/3}\ll 1$, not just when $(1-r_0/R_\star)\ll
1$, because acceleration is very slow in the planar solution.

The postshock and rarefaction flow are profoundly affected by
sphericity, but in a way that is consistent among different
progenitors. The effects that corrupt the planar solution are
primarily geometrical in nature -- that is, they occur because of
differences in the initial distribution (with scale length $R_\star$),
or differences in the hydrodynamics (also with scale length
$R_\star$). Insofar as the initial distributions of material are
identical after rescaling (i.e., so long as as
eq. [\ref{eq:outerform}] is correct), the fully spherical processes of
shock propagation and expansion in different progenitors are
identical. This implies that the outer ejecta distributions of
polytropes share \emph{common forms}, which limit to the self-similar
power law solutions at the highest velocities. Specifically, one can
multiply these power laws by a function that depends on the initial
depth, in order to account for the effects of progenitor structure and
spherical expansion.  The validity of this approach is demonstrated in
\S \ref{SS:commonform}, where such correction factors are also
presented.
 This approach is similar to the analytical solution of
Kazhdan \& Murzina (1992), who present a correction factor
(parameterized by $\gamma$ and $n$) that accounts for the first-order
effect of sphericity at all times in the flow. 
As shown in the Appendix, the correction factors we derive resemble the
outcome of Litvinova \& Nad\"ezhin's (1990) suggestion that spherical
effects truncate planar expansion once the radius has expanded by a
certain amount (of order $R_\star$).

To put it another way, it is possible to define a set of variables
that remove the scalings of physical quantities in the outer edge 
of a star. In terms of these variables, polytropes of the same index
assume the same structure (no longer up to a multiplicative constant)
at large initial radii. And, since the new time and velocity variables
have taken out variations in the strength of the shock as it enters
the outermost regions, different polytropic progenitors play out the
same behavior of shock acceleration and outer envelope ejection -- a
behavior that includes the effects of the finite radius on the initial
conditions and on the hydrodynamics of spherical expansion. Equation
(\ref{eq:outerform}) demonstrates that the normalized initial radius
$r_0/R_\star$ (or the normalized initial depth,
$x_0\equiv(R_\star-r_0)/R_\star$) and the rescaled mass coordinate
$(1-\m)/(\rho_1/\rho_\star)$ are variables of the required type, as is
the density variable $\rho/\rho_1$.  (The results of the Appendix, as
well as the analytical development of Kazhdan \& Murzina [1992], 
show that the variable $x_0^{1/3}$ is more relevant than $x_0$, and
so we define the corresponding mass variable $\hat{\delta m} \equiv
[(1-\m)/(\rho_1/\rho_\star)]^{1/3(n+1)}$.)  For the shock strength, we
must rely on the results of \S \ref{S:vshock} to generate a new
velocity variable $(v/v_\star) (\rho_1/\rho_\star) ^{\beta_1}$ that
removes the scaling of shock velocity with the structure of the
star. In terms of these variables, the final distribution of
high-velocity ejecta from polytropes of the same index $n$ share a
single distribution, which need only be determined once. Since it is
convenient to write this distribution as a power law (the self-similar
result) times a correction factor, it is natural to express the
correction factor in terms of $x_0^{1/3}$, $\hat{\delta m}$, or
$(v_f/v_\star)(\rho_1/\rho_\star)^{\beta_1}$, as is done in \S
\ref{SS:commonform}.

However, such correction factors must themselves fail to describe the
ejecta at some depth, or for velocities that are too low. This follows
from the fact that equation (\ref{eq:outerform}) is valid only in the
region for which the enclosed mass is essentially $M_\star$. For
radiative envelopes of constant opacity, the effective polytropic
index deviates from $3$ for the interior mass shells. For polytropes
as well, the initial density distribution is affected: the actual
density distribution (equation [\ref{eq:rhoform}]) differs from the
common distribution (equation [\ref{eq:outerform}]) when $m(r)+M_{\rm
rem}<M_\star$. Even when the shock wave is in the outermost mass
zones, its velocity is still being affected by the forward
characteristics of the earlier blastwave flow. At what point does the
flow forget its history and limit to a common distribution in these
outer regions? Theory (Whitham 1960) suggests that a shock's velocity
behavior is determined by local conditions once it begins to
accelerate strongly. The approach adopted in \S \ref{SS:commonform} is
to determine numerically what region of the ejecta is reproduced by a
variety of progenitors, and to present fits for correction factors
only in this region.

There is an upper velocity limit on the validity of our solutions as
well, because the finite width of the radiative shock becomes
comparable to the depth at some point. Our assumption of adiabatic
flow must break down by this point. We estimate the maximum
velocity in \S \ref{SS:vlimit}. 

\subsection{Limiting power laws for the highest-velocity ejecta}
\label{SS:ssresults} 
From our shock velocity approximation and the resulting entropy
distribution, equations (\ref{eq:vshock}) and (\ref{eq:sshock}), the
outer coefficients of shock velocity and entropy are, for $\gamma=4/3$,
\begin{eqnarray}\label{eq:fv1}
f_{v_s}(1) &=& \Gamma f_{\rho_0}(1)^{-\beta_1},\nonumber \\
f_s(1) &=& 0.1273 \Gamma^{3/2}f_{\rho_0}(1)^{-(3\beta_1/2+1/4)}. 
\end{eqnarray} 
In the planar, self-similar solution, $v_f(\m)/v_s(\m)$ is a constant,
$(v_f/v_s)_{\rm p}$, whose value we derive in the Appendix
and present in Table \ref{t:selfsim}.  This ratio gives $f_v(1)$ from
$f_{v_s}(1)$; then, mass continuity and entropy conservation give
$f_\rho(1)$ and $f_p(1)$:
\begin{eqnarray}\label{eq:fv1frho1fp1}
 f_v(1) &=& f_{v_s}(1) \left(\frac{v_f}{v_s}\right)_{\rm p}, \nonumber\\
 f_\rho(1) &=& \frac{1}{4\pi}\left(\frac{n+1}{\beta_1 n}\right)
	 f_{\rho_0}(1)^{3\beta_1}
	\left(\frac{v_f}{v_s}\right)_{\rm p}^{-3}, \nonumber \\
 f_p(1) &=& 0.002191 \left(\frac{n+1}{\beta_1 n}\right)^{4/3}
	\Gamma^{-2}f_{\rho_0}(1)^{2\beta_1-1/3}
	\left(\frac{v_f}{v_s}\right)_{\rm p}^{-4}. 
\end{eqnarray}
The exponents of these variables in $(1-\m)$ are easily derived, but
we state them here for convenience (the values are given in Table \ref{t:selfsim}): 
\begin{eqnarray}\label{eq:ws}
 w_v &=& w_{v_s} = -\beta_1 w_{\rho_0} = -\beta_1 \left(\frac{n}{n+1}\right),
 \nonumber\\
 w_s &=& \left(\frac{1}{\gamma}-1\right)w_{\rho_0} +
 \frac{2}{\gamma}w_v = \frac{n}{\gamma(n+1)}(1-2\beta_1-\gamma),
	\nonumber\\
 w_\rho &=& 1-3w_v = 1+3\beta_1\left(\frac{n}{n+1}\right),\nonumber\\
 w_p &=& \gamma(w_s+w_\rho) = \frac{\gamma + n - 2\beta_1 n + 3\beta_1
 \gamma n}{n+1}.
\nonumber\\
\end{eqnarray}

As an example, the asymptotic velocity distribution is:
\begin{eqnarray} \label{eq:vfouterform}
v_f(\m\rightarrow 1) &=& f_v(1) (1-\m)^{w_v}v_\star  \nonumber \\ &=& \Gamma
\left(\frac{v_f}{v_s}\right)_{\rm p} f_{\rho_0}(1)^{-\beta_1}
(1-\m)^{-\beta_1 n/(n+1)}v_\star.
\end{eqnarray} 

The asymptotic final density and pressure distributions are given in
terms of velocity by $\rho_f\propto v_f^{l_{\rho\,2}}t^{-3}$ and
$p_f\propto v_f^{l_{p\,2}}t^{-4}$, where
\begin{eqnarray}\label{eq:lrho2lp2}
l_{\rho\,2} &=& w_\rho/w_v = -\frac{n + 1 + 3\beta_1 n}{\beta_1 n},\nonumber\\ 
l_{p\,2} &=& w_p/w_v = -\frac{\gamma + n+ \beta_1 n(3\gamma-2)}{\beta_1
n}.
\end{eqnarray}
(The subscript 2 denotes power laws for the outer ejecta -- see \S
\ref{SS:harmRSGs}.) The values for $n=3/2$ and $n=3$ are given in
Tables \ref{t:harmonics} and \ref{t:radharms}, respectively. Based on
comparison with actual stellar models (\S \ref{S:simcomp}), there is 
some indication that even when the local value of $n_{\rm eff}\equiv d
\log\rho_0/d \log x_0$ varies, equation (\ref{eq:lrho2lp2}) still
roughly approximates $d\log\rho_f/d\log v_f$ and $d\log p_f/d\log v_f$
for the high-velocity ejecta of peripheral material. Note that a
shallower initial density produces a steeper final density and
pressure in the outermost ejecta.

\subsection{Common spherical dynamics of high-velocity ejection}
\label{SS:commonform} 

In Figures \ref{fig:vfvsfig} and \ref{fig:commonfig} we demonstrate
that convective and radiative stellar envelopes indeed each assume a 
common exterior distribution, as discussed above; that the shock velocity
approaches a common form for each (up to a constant); and that the
ratio $v_f(r_0)/v_s(r_0)$ does as well. In the Appendix we
investigate the implications of the suggestion of Litvinova \&
Nad\"ezhin (1990) that the final velocity is set by the truncation of
planar, self-similar acceleration once a shell's radius has expanded
to a radius somewhat larger than $R_\star$. This, along with the form
of the self-similar dynamics (equation [\ref{eq:vfvsfromS}]) gives a
form for $v_f/v_s$, equation (\ref{eq:vfvstrunc}) that is a polynomial
in $x_0^{1/3}$, where $x_0\equiv 1-r_0/R_\star$ is the normalized
depth. It is no surprise, then, that the numerically determined
distribution of $v_f(r_0)/v_s(r_0)$ is best fit and expressed as a
function of $x_0^{1/3}$ (as shown analytically by Kazhdan \& Murzina 1992):
\begin{eqnarray}\label{eq:vfvsUniversal}
\frac{v_f(x_0)}{v_s(x_0)} &\simeq&
\left(\frac{v_f}{v_s}\right)_{\rm p}
	(1 - 0.51x_0^{1/3} + 0.76x_0^{2/3} - 1.19x_0)
 \nonumber \,\,\,\,\,\,(n=3/2), \\
\frac{v_f(x_0)}{v_s(x_0)} &\simeq& 
\left(\frac{v_f}{v_s}\right)_{\rm p}
(1 - 0.34 x_0^{1/3} + 0.24 x_0^{2/3}  - 0.37 x_0)
\,\,\,\,\,\, (n=3),
\end{eqnarray}
where $(v_f/v_s)_{\rm p}$ is given in Table \ref{t:selfsim}.
These fits were made to the outer half of the radius of the $q=0.3$
polytrope in each case.  The fit for $n=3/2$ is good to within 1\% for
$r_0>0.55$ for both $q=0$ and $q=0.3$ and to within 3\% for $r_0>0.2$
for $q=1$ polytropic envelopes. The fit for $n=3$ is good to
within 1\% for $r_0>0.5$, $0.6$, and $0.65$ for $q=0.3$, $q=0$, and
$q=1$ respectively. As shown in Figure \ref{fig:vfvsfig}, 
the full spherical result, equation (\ref{eq:vfvsUniversal}),
resembles the outcome of Litvinova \& Nad\"ezhin's suggestion,
equation (\ref{eq:vfvstrunc}), if acceleration halts at about
$3R_\star$. 

We note that the first-order correction terms in equation
\ref{eq:vfvsUniversal} are predicted by the analytical perturbation
theory of Kazhdan \& Murzina (1992). From their theory, the
first-order coefficient should be $-0.3253$ for $n=3$ above, where we
find $-0.34$ in our fit. The two coefficients are somewhat different
in meaning, in that the Kazhdan \& Murzina coefficient represents a
term in the expansion of $v_f/v_s$ around $x_0^{1/3}=0$, whereas ours
is part of a cubic polynomial fit to a range of the variable
$x_0^{1/3}$.  The value $-0.3252$ for the first term in this expansion
implies that planar acceleration of the outermost material is
effectively truncated at $2.94 R_\star$ in Litvinova \& Nad\"ezhin's
scenario, very close to our estimate of $3R_\star$ (see the Appendix).

In the region for which envelopes of the same value of $n$ scale to a
common solution, we can give the Lagrangian form of the ejecta
distributions, i.e., $v_f(\m)$, $\rho_f(\m)t^3$ and $p_f(\m) t^4$.  We
present these high-velocity formulae in Table \ref{t:highv}. These
formulae consist of the power laws derived in \S \ref{SS:ssresults},
multiplied by a function of mass that accounts for the deviations from
self-similar flow. Since the mass external to a depth $x_0$ is
$1-\m(x_0) = 4\pi x_0^{n+1} (\rho_1/\rho_\star)/(n+1)$, the mass
variable corresponding to $x_0^{1/3}$ can be defined as: 
\begin{equation}\label{eq:dmhatdef}
\hat{\delta m}\equiv
[(1-\m)(\rho_1/\rho_\star)^{-1}]^{1/3(n+1)}. 
\end{equation} 
This variable is convenient for giving the correction factors to power
laws in $(1-\m)$, as in Table \ref{t:highv}, because the effects of
spherical expansion are linear in $\hat{\delta m}$ just as they are in
$x_0^{1/3}$. 

\subsection{Shock breakout and the maximum ejecta velocity}\label{SS:vlimit}

When a supernova shock approaches the surface of the star, there comes
a point at which the postshock radiation can leak out in a burst of
ionizing radiation (e.g., Chevalier \& Klein 1979). This event marks
the end of shock acceleration, and limits the velocity at which ejecta
can be cast away. The shock velocity at breakout is well-determined
from simple considerations. This sets the upper velocity limit for the
assumption of adiabatic flow, and the gross properties of the
radiation outburst (\cite{IN89}).

Shock propagation is unaffected by the leakage of radiation until the
optical depth of the shock exceeds the optical depth of the available
material. Since the shock wave is a region in which light diffuses
upstream and gets amplified by compression, the time for light to
diffuse across the shock front must be comparable to the time for
material to flow through it. This equality sets the physical size of
the shock transition: a simple estimate shows that the shock's optical
depth is
\begin{equation}\label{eq:shockwidth}
\tau_s \simeq \frac{c}{v_s}. 
\end{equation}
This is only an approximate equality, because the shock structure
set by radiation diffusion is not sharp. In fact, the structure of
a strong, radiation-dominated shock front propagating into a uniform
medium can be calculated analytically using the diffusion
approximation (Weaver 1976). In this solution, the radiation pressure
varies with optical depth as:
\begin{equation} \label{eq:shockptau}
\frac{p}{\rho_0 v_s^2} = \frac{6\exp{[-3(\tau-\tau_0)v_s/c]}}{1 +
7\exp{[-3(\tau-\tau_0)v_s/c]}}, 
\end{equation}
where $\tau$ is the optical depth (increasing inward) and $\tau_0$ is
a constant that defines the position of the shock front. Here, the
radiation pressure approaches its upstream and downstream asymptotes
exponentially. However, the smoothness of the shock front introduces
little uncertainty into the velocities ($\sim 20\%$) and
temperatures ($\sim 10\%$) derived below (\cite{IN89}). 

Shock breakout occurs when $\tau_s$ is equal to the stellar optical
depth $\tau(x_0)$ at some depth $x_0$:
\begin{equation}\label{eq:startau}
\tau(x_0) = \frac{\kappa \rho_1 R_\star x_0^{n+1}}{n+1}, 
\end{equation}
or, equivalently, when the radiation diffusion time across the shock
front equals the time for the shock to get to the stellar surface (the
local dynamical time).  The terminal shock velocity $v_s({\rm
breakout})$ at this point, where $\tau_s = \tau(x_0)$, is, from
equations (\ref{eq:vshock}), (\ref{eq:shockwidth}),
(\ref{eq:startau}):
\begin{equation}\label{eq:vfmaxanalytical}
v_{s}({\rm breakout}) \simeq \Gamma v_\star 
\left[
\left(\frac{\rho_1}{\rho_\star}\right)^{-1/n}
\left(\frac{\Gamma}{1+n}\right) 
\left(\frac{v_\star}{c}\right)
\left(\frac{\kappa M_{\rm ej}}{R_\star^2}\right)
	\right]^{\beta_1 n / (1+n-\beta_1 n)}. 
\end{equation}
The breakout shock velocity scales most significantly with $v_\star$,
because $\beta_1$ is a small number. 

It is interesting to note that the width of the shock at the time of
breakout is about $0.014R_\star$ and $0.019R_\star$ for RSGs and BSGs,
respectively. So, the parameter $x_0^{1/3}$ is at least $0.24$ and
$0.27$, respectively, for the adiabatic ejecta. Therefore the
planar, self-similar limit is never strictly achieved. 

\subsubsection{The speed limit for supernova ejecta} \label{SSS:speedlimit}

One might expect that after breakout, the diffusion of radiation would
interfere with adiabatic expansion at lower and lower velocities. If
this were true, even those mass shells not involved in the shock
at the time of breakout would be affected, and our adiabatic results
would not be applicable to these shells. 

However, a straightforward comparison of the diffusion time
$(\sqrt{3}\tau^2/c\kappa\rho)$ and the dynamical time $[t-t({\rm
breakout})]$ for a particular mass shell shows that this does not
actually happen. Instead, the adiabatic approximation holds for mass shells
deeper than a few shock widths from the surface at the time of
breakout.

To see this, consider the phase of planar free expansion, when a mass
shell has traveled far compared to its initial depth, but only a small
distance relative to $R_\star$. Then, because the surface area of a
shell is effectively constant, the density scales as $\rho(m)\propto
1/[t-t({\rm breakout})]$. Since the optical depth $\tau$ of a mass
shell does not vary in planar flow, the diffusion time is proportional
to the dynamical time during this planar free expansion
phase. Examining the planar, self-similar solution presented in the
Appendix, we find that the ratio of the local diffusion time to
dynamical time is never less than its value in free
expansion. Similarly, the shock front itself has a width that is set
by the equality of the diffusion time across the front to the time for
material to cross the shock front. As a result, the criterion for a
mass shell to experience significant radiation diffusion during planar
expansion is simply that it be entrained in the shock at the time of
breakout.

Even so, one might still worry that the onset of spherical expansion
will permit a diffusion wave to propagate inward. Then, the geometric
reduction of a mass shell's velocity would be enhanced (and the
formulae of \S \ref{SS:commonform} invalidated) by the loss of
photons. Fortunately, this does not happen either. The results of \S
\ref{SS:commonform} indicate that the geometric effect sets in at the
fixed radius $3R_\star$. Expansion to a radius of $3R_\star$ decreases
the ratio of diffusion time to dynamical time by only a factor of
about $\sim 9$ relative to its planar value. One only has to move to
slightly ($\sim 30\%$) lower velocity to recover this factor between
the two time scales. Therefore, geometrical expansion ends adiabatic
acceleration before radiation diffusion ever becomes important --
except for the material caught in or very near the shock front at the
moment of emergence. (See Chevalier 1992 for an analysis of a diffusion
wave in spherical free expansion.)

So, we are justified in applying the full, adiabatic, planar
acceleration factor $(v_f/v_s)_{\rm p}$ to the shock velocity at
breakout in order to estimate the upper velocity limit $v_{f\,{\rm
max}}$ for which adiabatic results hold. 

For convective (n=3/2) and radiative (n=3) envelopes of typical RSG
and BSG dimensions, respectively, we find: 
\begin{eqnarray}\label{eq:vfmax}
v_{f\,{\rm max}} &=& 13,000
 \left(\frac{\kappa}{0.34\, {\rm cm}^2\, {\rm
g}^{-1}}\right)^{0.13} 
\left(\frac{\rho_1}{\rho_\star}\right)^{-0.086}
\nonumber\\
&\times&\left(\frac{E_{\rm in}}{10^{51} {\rm erg}}\right)^{0.57}
\left(\frac{M_{\rm ej}}{10 M_\odot}\right)^{-0.44}
\left(\frac{R_\star}{500 R_\odot}\right)^{-0.26}
{\rm km\,s^{-1}}\;\;\;\;\;({\rm n=3/2}),\nonumber\\ 
v_{f\,{\rm max}} &=& 33,000
\left(\frac{\kappa}{0.34\,{\rm cm}^2\,{\rm g}^{-1}}\right)^{0.16} 
\left(\frac{\rho_1}{\rho_\star}\right)^{-0.054}
\nonumber\\
&\times& \left(\frac{E_{\rm in}}{10^{51} {\rm erg}}\right)^{0.58}
\left(\frac{M_{\rm ej}}{10 M_\odot}\right)^{-0.42}
\left(\frac{R_\star}{50 R_\odot}\right)^{-0.32}
{\rm km\,s^{-1}}\;\;\;\;\;\;\;\;({\rm n=3}).\nonumber\\ 
\end{eqnarray}
Note the insensitivity of $v_{f\,{\rm max}}$ to the
structure parameter $\rho_1/\rho_\star$.

\subsubsection{Relativistic mass ejection}\label{SSS:relativity}

Recently, Kulkarni et al. (1998) have inferred that the radio-bright
shell around the peculiar supernova 1998bw expanded relativistically
during the period of observation, and take this to be physical
evidence of the possible association (Soffita et al. 1998) with the
gamma-ray burst GRB 9804025. Woosley, Eastman \& Schmidt (1998)
considered models for 1998bw with ejected masses in the range 5-12
$M_\odot$ and energies in the range 4--30$\times 10^{51}$ ergs. We are
motivated by these developments to reconsider the question of
relativistic mass ejection, addressed initially by Colgate \& Johnson
(1960) and Colgate \& White (1966) and most recently by Woosley et al.
(1998).

Our criterion for relativistic mass ejection is that the maximum
ejecta velocity given by our non-relativistic theory, equation
(\ref{eq:vfmax}), exceed the speed of light. (For the behavior of
shock waves in the relativistic regime, see \cite{CJ60} and
\cite{JM71}). The star must be sufficiently compact to meet this
criterion. For instance, applying equation (\ref{eq:vfmax}) to an
explosion of $10^{51}$ ergs in a white dwarf, $v_{f\,{\rm max}}$ would
be of order $1.7c$ if $n=3/2$, climbing to $\sim 5.4c$ if $n=3.25$ in
the atmosphere. But, the helium star progenitors of ordinary type Ib
and Ic supernovae may not be compact enough.  Taking stellar and
ejected masses for such progenitors as considered by Woosley, Langer
\& Weaver (1995), and using their radii at carbon ignition (1--9
$R_\odot$) and an effective outer polytropic index $n_{\rm eff}=3$ to
represent the presupernova star, explosions of $10^{51}$ ergs produce
maximum final velocities in the range $0.5c$--$0.7c$.

In the periphery of a star, the shock velocity is a simple power law
of initial density (equation [\ref{eq:vshock}]), with the coefficient
given by equation (\ref{eq:Gamma}). In order for the non-relativistic
extrapolation of the final velocity of a mass shell to exceed $c$, its
initial density must be smaller than the mean density by a critical
factor: taking $\beta_1=0.19$ and $v_f/v_s = 2.1$ in equation
(\ref{eq:vshock}),
\begin{equation}\label{eq:rhoRelativistic}
\rho_0 \lesssim \left(\frac{1.7 v_\star}{c}\right)^{5.3} \rho_\star.
\end{equation}
But, the electron-scattering optical depth of this
material must exceed about $2$, or the radiation driving the shock will
escape. For an outer density law $\rho_0=\rho_1 (1-r_0/ R_\star)^n$,
this criterion puts an upper bound on the radius: 
\begin{equation}\label{eq:Rrelativistic}
R_\star \lesssim 10^{13.9-2.4/n} \left[\frac{\kappa/0.34\,
{\rm cm^2\, g^{-1}}}{(1+n)(\rho_1/\rho_\star)^{1/n}}\right]^{1/2}
\left(\frac{E_{\rm in}}{10^{52} {\rm erg}}\right)^{1.3(1+1/n)}
\left(\frac{M_{\rm ej}}{1 M_\odot}\right)^{1/2-1.3(1+1/n)} {\rm cm}.
\end{equation}
If the star is sufficiently compact by the above criterion, equations
(\ref{eq:rho1f1}) and (\ref{eq:vfouterform}) give a rough estimate for
the mass of relativistic ejecta:
\begin{equation}\label{eq:mRelativistic}
M_{\rm rel}\sim \frac{10^{29.6-4.8/n}}{n+1}
\left(\frac{\rho_1}{\rho_\star}\right)^{-1/n}
\left(\frac{E_{\rm in}}{10^{52} {\rm erg}}\right)^{2.6(1+1/n)}
\left(\frac{M_{\rm ej}}{1 M_\odot}\right)^{1-2.6(1+1/n)} {\rm g}.
\end{equation}
We have assumed that the initial depth of this material is small
compared to $R_\star$. An explosion of $10^{52}$ ergs ejecting
$1M_\odot$ in a sufficiently compact star (with $n=3$ or, as
considered by Colgate \& Johnson 1960, $n=3.25$) will produce
relativistic ejecta with $M_{\rm rel}c^2\simeq 3\times 10^{48}$ ergs,
for typical values of $\rho_1/\rho_\star$. (But, to compare, we would
predict a shock velocity $50\%$ higher than Colgate \& Johnson found,
at the point where $\rho_0=30\,{\rm g\,cm^{-3}}$ in their progenitor.)
This energy is comparable to the minimum energy associated with
relativistic electrons in SN1998bw ($10^{49}$ ergs; Kulkarni et
al. 1998), and greater than the energy in the gamma-ray burst
($8.5\times 10^{47}$ ergs; Galama et al. 1998), if the burst and
supernova are associated.

\subsubsection{Observable aspects of shock emergence}
\label{SSS:emergence}

The shock velocity approximation of \S \ref{S:vshock} allows us to
predict the variation of the breakout quantities with the explosion
parameters $E_{\rm in}$, $R_\star$, $M_\star$ and $\rho_1/\rho_\star
$. The quantities that can be extracted from simple analytic
calculations include the postshock radiation temperature $T_{\rm se}$,
the outburst energy $E_{\rm se}$, and the time scale on which
radiation diffuses out of the shock, $t_{\rm se}$ (assuming
non-relativistic ejection).  Detailed information about shock
emergence, like the hydrodynamics of optically thin material (which
has been suggested to form a thin, dense shell, e.g., \cite{EB92}) or
the spectrum of the outburst, are beyond the scope of our simple
analysis. The outburst spectrum is affected by the work that photons
do as they diffuse out of the shock; a detailed solution awaits
multifrequency radiation hydrodynamics simulations. Nonetheless, the
values of $T_{\rm se}$ given below agree with recent estimates of the
peak color temperature in outburst (e.g., \cite{EB92}) within 10\%.

The postshock radiation temperature is related to the postshock 
pressure at the time of shock emergence, so $a T_{\rm se}^4/3 = 2\rho_0
v_s^2({\rm breakout})/(\gamma+1)$: 
\begin{eqnarray}\label{eq:Tse}
T_{\rm se} &=& 
5.55 \times 10^5 
\left(\frac{\kappa}{0.34\,{\rm cm}^2\,{\rm g}^{-1}}\right)^{-0.10}
 \left(\frac{\rho_1}{\rho_\star}\right)^{0.070} 
\nonumber\\ &\times&
 \left(\frac{E_{\rm in}}{10^{51} {\rm erg}}\right)^{0.20}
\left(\frac{M_{\rm ej}}{10 M_\odot}\right)^{-0.052}
\left(\frac{R_\star}{500 R_\odot}\right)^{-0.54}
{\rm K}\;\;\;\;\;\;\;\;({\rm n=3/2}),\nonumber\\ 
T_{\rm se} &=& 
1.31 \times 10^6
\left(\frac{\kappa}{0.34\,{\rm cm}^2\,{\rm g}^{-1}}\right)^{-0.14}
 \left(\frac{\rho_1}{\rho_\star}\right)^{0.046}
\nonumber\\ &\times&
 \left(\frac{E_{\rm in}}{10^{51} {\rm erg}}\right)^{0.18}
\left(\frac{M_{\rm ej}}{10 M_\odot}\right)^{- 0.068}
\left(\frac{R_\star}{50 R_\odot}\right)^{- 0.48}
{\rm K}\;\;\;\;\;\;\;\;({\rm n=3}).
\end{eqnarray}

The energy $E_{\rm se}$ of the radiation outburst can be estimated as
the thermal energy in the shock front at the time of breakout. So,
$E_{\rm se} \simeq (a T_{\rm se}^4/3)[4\pi R_\star^3 x_0({\rm
breakout})]$: 
\begin{eqnarray}\label{eq:Ese}
E_{\rm se} &=& 
1.7 \times 10^{48} 
\left(\frac{\kappa}{0.34\,{\rm cm}^2\,{\rm g}^{-1}}\right)^{-0.87}
 \left(\frac{\rho_1}{\rho_\star}\right)^{- 0.086} 
\nonumber\\ &\times&
 \left(\frac{E_{\rm in}}{10^{51} {\rm erg}}\right)^{0.56}
\left(\frac{M_{\rm ej}}{10 M_\odot}\right)^{- 0.44}
\left(\frac{R_\star}{500 R_\odot}\right)^{1.74}
{\rm ergs}\;\;\;\;\;\;\;\;({\rm n=3/2}),\nonumber\\ 
E_{\rm se} &=& 
7.6 \times 10^{46}
\left(\frac{\kappa}{0.34\,{\rm cm}^2\,{\rm g}^{-1}}\right)^{-0.84}
 \left(\frac{\rho_1}{\rho_\star}\right)^{- 0.054}
\nonumber\\ &\times&
 \left(\frac{E_{\rm in}}{10^{51} {\rm erg}}\right)^{0.58}
\left(\frac{M_{\rm ej}}{10 M_\odot}\right)^{- 0.42}
\left(\frac{R_\star}{50 R_\odot}\right)^{1.68}
{\rm ergs}\;\;\;\;\;\;\;\;({\rm n=3}).
\end{eqnarray}

The energy $E_{\rm se}$ will be released on the diffusion time at
shock breakout, which is also the time for the shock to travel its
width. So, 
\begin{eqnarray} \label{eq:tse}
t_{\rm se} &=& 790
\left(\frac{\kappa}{0.34\,{\rm cm}^2\,{\rm g}^{-1}}\right)^{- 0.58}
 \left(\frac{\rho_1}{\rho_\star}\right)^{- 0.28} 
\nonumber\\ &\times&
 \left(\frac{E_{\rm in}}{10^{51} {\rm erg}}\right)^{- 0.79}
\left(\frac{M_{\rm ej}}{10 M_\odot}\right)^{ 0.21}
\left(\frac{R_\star}{500 R_\odot}\right)^{2.16}
{\rm sec}\;\;\;\;\;\;\;\;({\rm n=3/2}),\nonumber\\ 
t_{\rm se} &=& 40
\left(\frac{\kappa}{0.34\,{\rm cm}^2\,{\rm g}^{-1}}\right)^{-0.45}
 \left(\frac{\rho_1}{\rho_\star}\right)^{- 0.18}
\nonumber\\ &\times&
 \left(\frac{E_{\rm in}}{10^{51} {\rm erg}}\right)^{- 0.72}
\left(\frac{M_{\rm ej}}{10 M_\odot}\right)^{ 0.27}
\left(\frac{R_\star}{50 R_\odot}\right)^{1.90}
{\rm sec}\;\;\;\;\;\;\;\;({\rm n=3}).
\end{eqnarray}
However, note that the pulse will be longer than this from the vantage
of a distant observer because of the light travel time (e.g.,
\cite{EB92}), which is $1160$ seconds for a star of radius $500
R_\odot$, or $116$ seconds for a star of radius $50 R_\odot$. As a
result, the observed luminosity will be typically somewhat less than
$E_{\rm se}/t_{\rm se}$. 

These formulae reproduce the energy and time scale of the radiation
outburst in the numerical simulations of Ensman \& Burrows (1992), for
explosion parameters chosen to match theirs.  These formulae are also
in agreement with the analytical results of Imshennik \& Nad\"ezhin
(1989) for SN1987A, but only if we account for the fact that the shock
velocity they use is $47\%$ faster than is given by equation
(\ref{eq:vshock}) for the same model, as if $\Gamma$ were $1.16$
instead of $0.79$ for the outermost mass shells. If instead we use
equations (\ref{eq:Tse}) and (\ref{eq:Ese}), we find a value of
$T_{\rm se}$ that is $20\%$ lower, and a value of $E_{\rm se}$ that is
$50\%$ lower, than those of Imshennik \& Nad\"ezhin. It is not
apparent why the two shock formulae do not agree, as their formula is
taken from their simulations, and our formula agrees with our
simulations.

For the helium star progenitors of typical type Ib and Ic supernovae,
described by Woosley, Langer and Weaver 1995 and discussed in \S
\ref{SSS:relativity} (again, assuming that $E_{\rm in}=10^{51}$ ergs),
the energy associated with the outburst would be between $3\times
10^{44}$ and $2\times 10^{46}$ ergs, and the outburst would last
between two and twenty seconds.  In making these estimates, we have
assumed that the stellar wind does not have sufficient optical depth
to support a radiation-dominated shock.  For compact progenitors with
dense Wolf-Rayet winds, this assumption may fail.  In that situation,
the shock velocity formula (equation \ref{eq:vshock}) can be applied
to the stellar wind, and breakout quantities estimated at the radius
for which the wind optical depth matches $\sim c/v_s$. The
circumstellar interaction (e.g., Fransson, Lundqvist \& Chevalier
1996) would then begin immediately.

\section{Pressure-based model for the ejecta pressure and density
distributions} 
\label{S:pressuremodels}  

We now wish to present a model that describes the distribution of all
the ejecta, not just the high velocity ejecta discussed in \S
\ref{S:highv}. To do so, we will extend the form of the
\emph{pressure} distribution inward in mass, approximating its
variation with a simple functional form. In our calculations, the
final pressure distribution is invariably the smoothest of the
hydrodynamical variables (expressed as functions of $\m$); this is not
surprising, since the pressure gradient is inhibited on small scales
by its ability to accelerate the material so as to reduce its
magnitude. Once we make a model for the final pressure distribution,
the final density distribution follows immediately from the entropy
left behind by the forward shock.

To make a simple model for the final pressure distribution, $p_f(\m)
t^{4}$, we multiply the pressure distribution known for the
high-velocity material, given in Table (\ref{t:highv}), by
a simple function of $m$. This has the advantage of preserving the
high-velocity behavior of the flow derived in \S \ref{S:highv}, because
$\m$ varies very little in the region of validity of this
solution. The solution found in this manner is less accurate than the
solutions of \S \ref{S:highv}, but they have the advantage that they are
robust to variations in the progenitor structure (like the existence
of superadiabatic gradients in RSGs). They fail in regions that
experience a strong reverse shock, especially the mantles of RSGs. 
However, as we discuss in \S \ref{SS:rhojump}, the density jump between
the mantle and outer envelope ejecta \emph{can} be predicted despite
the formation of reverse shocks. 

Multiplying the pressure distribution specified in Table
\ref{t:highv} by a simple function of mass produces our model for the
final radiation pressure distribution: 
\begin{eqnarray}\label{eq:pmodel}
\frac{p_f(\m)t^{4}}{p_\star t_\star^{4}}
 &=& \left[\frac{p_f(\m)t^{4}}{p_\star t_\star^{4}}\right]_{\rm PL+HV}
\left[1 - \alpha(1-\m)\right]^{4/3},
\nonumber\\
\end{eqnarray}
where $\left[p_f(\m)t^{4}/p_\star t_\star^{4}\right]_{\rm PL+HV}$ is
the pressure distribution given in Table \ref{t:highv}, that is, a
self-similar outer power law times a correction factor that accounts
for the geometrical effects spherical expansion. The new term, $[1 -
\alpha(1-\m)]^{4/3}$, is but another correction factor that accounts
for our imperfect knowledge of the pressure distribution in the bulk
of the ejecta; we have chosen the power $4/3$ to simplify the result
for the final density distribution.  It only remains to specify the
free parameter $\alpha$ used in formulae (\ref{eq:pmodel}),
(\ref{eq:rhomodel}) and (\ref{eq:vmodel}). In \S \ref{S:RSGs} we will
develop a technique for generating realistic RSG progenitors with
which to constrain $\alpha$; the variation of $\alpha$ with the stellar
structure parameters of red supergiants is presented in Table
\ref{t:harmonics}.

The final density distribution corresponding to the ejecta pressure
model of equation (\ref{eq:pmodel}) is obtained by means of the
entropy deposited in the forward shock: 
\begin{equation}\label{eq:rhomodel1}
\frac{\rho_f(\m) t^3}{\rho_\star t_\star^3} =
\left[\frac{p_f(\m)t^{4}}{p_\star t_\star^{4}}\right]^{1/\gamma} 
\left[\frac{s(\m)}{s_\star}\right]^{-1},
\end{equation}
where $s(\m)$ is given by equation (\ref{eq:sshock}). Writing out
the resulting density model in terms of the variables $\m$, $r_0$ and
$\rho_0$, we find
\begin{eqnarray}\label{eq:rhomodel}
\frac{\rho_f(\m) t^3}{\rho_\star t_\star^3} &=&
\Lambda_{\rho} (1-\m)^{3w_p/4}
\m^{(3-6\beta_1)/4}
\left(\frac{r_0}{R_\star}\right)^{9\beta_1/2}
\left(\frac{\rho_0}{\rho_\star}\right)^{(1+6\beta_1)/4}
\nonumber \\
&\times&  \left[1-\alpha(1-\m)\right]
\left\{\begin{array}{lc} 
 (1+0.96\dmhat)^{3/4} & (n=3/2),  \\ 
 (1+1.03\dmhat+0.37\dmhat^2)^{3/4} &  (n=3).
\end{array} \right.
\end{eqnarray}

The only parameter remaining to be determined is the numerical
prefactor $\Lambda_\rho$. The planar, self-similar solution gives the
definite value $\Lambda_\rho=11 f_p(1)^{3/4}$, where $f_p(1)$ is given by equation
(\ref{eq:fv1frho1fp1}). However, we find that using overall energy
conservation to set $\Lambda_\rho$ gives more accurate results. To do
this, we first derive the velocity distribution from the density model
by means of a numerical integration:
\begin{equation}\label{eq:vmodel}
\frac{v_f(\m)^3}{v_\star^3} = \frac{3}{4\pi} \int_0^{\m}
\left[\frac{\rho_f(\m')t^3}{\rho_\star t_\star^3}\right]^{-1} d\m'. 
\end{equation}
(Note that by integrating from zero velocity, we have excluded the
remote possibility of hollow blastwaves, which can occur in special
distributions; see Ostriker \& McKee 1988.)
If we define the model energy $E_{\rm model}$ by,
\begin{equation}\label{eq:emodel}
\frac{E_{\rm model}}{E_{\rm in}} \equiv \frac{1}{2}\int_0^1
\frac{v_f(\m)^2}{v_\star^2} d \m',
\end{equation}
then the energy of the ejecta model can be brought into agreement with
$E_{\rm in}$ by the transformation (from eq. \ref{eq:scales}): 
\begin{eqnarray}\label{eq:energyconservation}
\frac{v_f(\m)}{v_\star} &\rightarrow& \frac{v_f(\m)}{v_\star}
 \left(\frac{E_{\rm model}}{E_{\rm in}}\right)^{-1/2}, \nonumber\\
\frac{\rho_f(\m) t^3}{\rho_\star t_\star^3} &\rightarrow&
\frac{\rho_f(\m) t^3}{\rho_\star t_\star^3}
 \left(\frac{E_{\rm model}}{E_{\rm in}}\right)^{3/2}, \nonumber\\
\frac{p_f(\m) t^{4}}{\rho_\star t_\star^{4}}&\rightarrow&
\frac{p_f(\m) t^{4}}{\rho_\star t_\star^{4}}
 \left(\frac{E_{\rm model}}{E_{\rm in}}\right)^{2}. 
\end{eqnarray}
In this transformation, the characteristic scales ($v_\star$, etc.)
are kept fixed but the model ejecta distributions ($v_f(\m)$, etc.) are
altered to agree with the blastwave energy. 

Equation (\ref{eq:rhomodel}) constitutes a local model for the final
ejecta distribution: one that specifies the final density of a mass
shell by reference to its initial variables $(m, r_0, \rho_0)$
only. The enforcement of overall energy conservation, and the
application of the parameter $\alpha$, weaken this statement only
slightly. This model inherits from the entropy and shock-velocity
models its inability to describe regions that have been hit by a
strong reverse shock. An underestimate of the entropy leads to an
overestimate of the density in the mantles of supergiants, RSGs
especially. 
           
\subsection{Density variation across the outer composition
boundary}\label{SS:rhojump} 
               
Independently of our model for the distribution of the outer envelope
ejecta, our shock velocity model allows us to evaluate the relative
jump in density at the base of the outer envelope. This boundary is
typically the outermost composition boundary in the progenitor. If we
take $p_f(\m) t^{4}$ to be continuous in $\m$ across the boundary, the
entropy left behind by the shock wave, equation (\ref{eq:rhomodel}) with
$\beta_1=0.19$, dictates:
\begin{equation}\label{eq:rhojump}
\delta\log(\rho_f t^3) =
0.54\delta\log(\rho_0)+0.86\delta\log(r_0), 
\end{equation}
where $\delta\log(\rho_0)$ and $\delta\log(r_0)$ are to be evaluated
for nearby mass shells straddling the composition boundary.

One might be concerned that this result would be invalidated by the
formation of a reverse shock at this boundary, or by the development
of the Rayleigh-Taylor instability. However, in our simulations the
compression wave that decelerates the mantle material travels inward
(in mass) before it steepens into a shock wave, so that the entropy
distribution given by equation (\ref{eq:sshock}) is correct across the
boundary. The Rayleigh-Taylor instability may indeed break the
spherical symmetry of the ejecta. However, it is unlikely to change
the fact that the outermost mantle ejecta is in contact with the
innermost ejecta of the outer envelope.  Then, pressure equilibrium
across the boundary gives the density jump above. An interesting
corollary is that the temperature distribution is continuous as well,
even once radiation diffusion becomes significant. This may have
implications for the recombination epoch, since helium (with a much
lower specific entropy) underlies any hydrogen-rich ejecta.

Those interested in modeling the distribution and expansion velocity
of the radioactive inner mantle material should note that our model
for the outer envelope ejecta constrains these quantities rather
poorly, because we overestimate the density of the mantle material.
 
\section{Harmonic mean models for ejecta distributions}\label{SS:harmRSGs}

Our simulations demonstrate that the final density and pressure in the
hydrogen ejecta can often be fit reasonably by the simple density model,
\begin{equation}\label{eq:harmrho}
\frac{\rho_f t^3}{\rho_\star t_\star^3} =
 \frac{\rho_{\rm break} t^3}{\rho_\star t_\star^3}
 \left[\frac{ 
	 \left(\frac{v_f}{v_{\rho\,{\rm break}}}\right)^{-l_{\rho\,1}/y_\rho}
	+\left(\frac{v_f}{v_{\rho\,{\rm break}}}\right)^{-l_{\rho\,2}/y_\rho}
	}{2}\right]^{-y_\rho}, 
\end{equation}
and the corresponding pressure model, 
\begin{equation}\label{eq:harmp}
\frac{p_f t^4}{p_\star t_\star^4} =
 \frac{p_{\rm break} t^4}{p_\star t_\star^4}
 \left[\frac{ 
	 \left(\frac{v_f}{v_{p\,{\rm break}}}\right)^{-l_{p\,1}/y_p}
	+\left(\frac{v_f}{v_{p\,{\rm break}}}\right)^{-l_{p\,2}/y_p}
	}{2}\right]^{-y_p}. 
\end{equation}
These models interpolate between limiting power laws at low and high
velocities, joined at a specified value ($\rho_{\rm break}$ or $p_{\rm
break}$) at a given break velocity ($v_{\rho\,{\rm break}}$ or 
$v_{p\,{\rm break}}$), with a curvature set by the parameters $y_\rho$
or $y_p$.

For supernovae in red supergiants, these models apply only to the
hydrogen ejecta, because of the severe density jump at the base of
this material. As a result, we cannot apply the constraints of mass
and energy conservation to give relations between the five parameters
$\rho_b t^3$, $v_{\rho\,{\rm break}}$, $l_{\rho\,1}$, $l_{\rho\,2}$
and $y_\rho$ in the density model.  Instead, we fix $l_{\rho\,2}$ and
$l_{p\,2}$ at their self-similar values; we set $y_\rho$ and $y_p$ at
characteristic values, and we fit the other three as functions of $q$,
using the techniques of \S \ref{S:RSGs}. Table \ref{t:harmonics}
presents these fits for red supergiants. 

These simple models fit the outer envelope ejecta of polytropic models
for red supergiants quite accurately, to about $9-11\%$ error in
density and $23-30\%$ error in pressure (both errors increasing with
increasing $q$). In fact, the residuals have a very consistent pattern
in each case, indicating that more complicated models could achieve
even greater accuracy. However, real stars may not resemble polytropes
sufficiently to justify any more complexity.

For supernovae in stars with radiative outer envelopes, such as blue
supergiants, no sequence of simple models models is available for the
fitting of these parameters. However, the difference between mantle
and outer envelope is not as severe for these stars, so we can apply
models to the entire ejecta at once. The constraints of mass and
energy conservation then reduce the number of parameters in the
density model from five to three. The outer power law behavior is
strongly constrained by the planar, self-similar solution. Chevalier
\& Soker (1989) conjectured that the interior power law could be
described by self-similar rarefaction, a hypothesis we also adopt to
set $l_{\rho\,1}$ and $l_{p\,1}$. Taking $y_\rho$ to be the same as in
the RSG models, the density distribution is fully specified. For the
parameters $y_p$, $v_{p\,{\rm break}}$ and $p_{\rm break}$ of the model
pressure, few constraints are available. We quote typical values
only. The result is a fiducial model, given in Table \ref{t:radharms},
that may be representative, but which depends not at all on the actual
progenitor structure. In this sense, the pressure-based model of \S
\ref{S:pressuremodels} is superior, especially for radiative
progenitors. See Table \ref{t:errors} for a synopsis of the relative
accuracy of these models for realistic blue and red supergiant
progenitors. 

\section{Survey of ejecta distributions from model red supergiants}
\label{S:RSGs}
The extreme difference in density between the mantles and outer
envelopes of red supergiants, as well as the relative constancy of the
hydrogen envelope density, suggest that simple models can reproduce
the dynamics and final density distribution of these stars. Such an
approach was taken by Chevalier (1976), who found that progenitor
models with constant-density mantles and constant-density outer
envelopes, separated by a region of steep density decline, could
reproduce the qualitative features of type II light curves. Most
notably, he identified existence of a plateau phase with the formation
of a `detached shell' (density maximum) in the final distribution. We
will use a different set of simplified models to represent these
progenitors, so it is worthwhile to review in detail why such models
are realistic.

In red supergiants, the outer convective envelope can be modeled as a
polytrope of index $n=3/2$ (\S \ref{S:progenitors}), so that the shape
of its density distribution is determined by the ratio of its mass to
that of the star, $q = 1-M_{\rm env}/M_{\rm ej}$. Clearly, the
initial distribution of density in the progenitor will affect the
final distribution of ejecta in the ejecta. 

There is one other parameter that plays an important role in
determining the ejecta distributions from the convective outer
envelopes of red supergiants: $M_{\rm env}/M_{\rm ej}$, the ratio of
the outer envelope's mass to the total ejected mass. This ratio
determines how much material (beneath the outer envelope) participates
in the explosion. Although the outer envelope will wind up with the
lion's share of the final kinetic energy, the mass of underlying
ejected mantle is important in the blastwave stage. At the beginning
of this phase, the explosion energy is contained within the mantle.
During the course of the blastwave, as the explosion energy gets
transferred to the outer envelope, the mass of the swept-up mantle
material continues to affect the forward shock velocity
(eq. \ref{eq:vshock}). 

Note that the shock velocity in the outer envelope (as approximated by
equation [\ref{S:vshock}]) is affected by the \emph{mass} of the
mantle, but not (to the accuracy of this approximation) by its
structure. Indeed, there is reason to believe that this property
should hold for the final ejecta distribution as well.  Our
pressure-based model for the ejecta is a local theory, making
reference to the initial variables $(m,r_0,\rho_0)$ in the
progenitor. The distinction between remnant and mantle affects the
mass coordinate $m$, so the mass ratio $M_{\rm env}/M_{\rm ej}$ affects our
prediction for the outer envelope ejecta. But, the initial radius and
structure of the mantle do not. So, we are justified in using
simplified models whose mantles have a realistic mass but an idealized
structure.

The two parameters $q=1-M_{\rm env}/M_\star$ and $M_{\rm env}/M_{\rm
ej}$ are clearly correlated for real progenitors, because $M_\star$
and $M_{\rm ej}$ only differ by the remnant mass $M_{\rm
rem}$. Indeed, only certain values of these two parameters are
realistic, as depicted by Figure \ref{fig:paramfig}. Although the loss
of the outer envelope mass in isolated stars and binaries is uncertain
and intrinsically variable, the evolution of stellar cores and mantles
appears to be nearly independent of mass loss and initial metallicity
while a hydrogen envelope persists (\cite{WW95}, \cite{M92}). Assuming
that the fiducial estimate of the remnant mass is correct, this
implies that $M_{\rm rem}$ and $M_{\rm man}$ are well known as a
function of initial stellar mass. The final value of the hydrogen
envelope mass $M_{\rm env}$ depends on uncertain mass loss, but it is
unlikely that the final stellar mass $M_\star$ of an RSG will exceed
the initial main-sequence mass, except perhaps in some massive
binaries. The circles in Figure \ref{fig:paramfig} demonstrate the
locus of the two parameters for stars that lose no mass; the stars on
this line are the models of Maeder (1992), which also agree with those
of Woosley \& Weaver (1995) and show no significant variation with
metallicity. Varying $M_{\rm env}$ between this maximum amount and
zero, a star with a certain ratio $M_{\rm rem}/M_{\rm man}$ will trace
out a curve (shown as dotted lines) in the space of these
parameters. As discussed by Woosley \& Weaver (1995), the mass loss
rate of isolated stars increases with metallicity. Their calculated
final states of solar-metallicity stars are shown as stars in the
figure.

It is evident from this figure that only a narrow swath of the
parameter space of $q$ and $M_{\rm env}/M_{\rm ej}$ is likely to be
inhabited by type II supernova progenitors. So, we are justified in
focusing our attention on the one-parameter family (dashed line
in Figure \ref{fig:paramfig}),
\begin{equation}\label{eq:qline}
\frac{M_{\rm env}}{M_{\rm ej}} = 1.12 - 1.08q, 
\end{equation}
for values of $q$ between 0.3 and 0.8. For comparison, we have also
considered values of $M_{\rm env}/M_{\rm ej}$ greater by $0.05$ at a
given $q$, in order to capture the variation of quantities over the
physical range.

Our model stars have a dense central region that represents the
mantle, which has a small radius (typically $10^{-4} R_\star$), a
uniform density, and the appropriate mass as given by $M_{\rm
man}/M_{\rm ej}$. Above this lies a polytrope of index $n=3/2$ that
represents the stellar envelope; its structure is given by the value
of $q$ under consideration. Explosions in these progenitors are then
simulated numerically as outlined in \S \ref{S:simcomp}, and the
results are analyzed to find the variation of the parameter $\alpha$
for the model of section \ref{S:pressuremodels} as a function of $q$. 

We base our trust in these models on demonstrations that their outer
envelope ejecta are insensitive to the initial mantle distribution and
can match the results of more sophisticated progenitor models. In
Figure \ref{fig:polyrsgfig}, we show that varying the mantle radius
(and thus its density) alters the final distribution of mantle
material slightly, but \emph{not} the final distribution of the outer
envelope material. This figure also shows that our models can match
the outer envelope ejecta in the explosion of the $15 M_\odot$ model
of Woosley, implying that large differences in the initial mantle
distribution lead to negligible differences in the final outer
envelope ejecta.
 
\section{Calibration of models against simulations}\label{S:simcomp}

\subsection{Simulation technique}
Our simulations are performed with an explicit, second-order,
Lagrangian, finite-differencing code, as described by Richtmeyer \&
Morton (1967). Our code was written by Kelly Truelove, modeled after a
code by John Holliman, as described in Truelove \& McKee (1998). We have
modified it slightly for the simulation of supernovae. Our progenitor
distributions were kindly supplied by Ken'ichi Nomoto and Stan
Woosley, and have been interpolated onto finer Lagrangian grids for
more accurate dynamical results. For the study of lone polytropic
envelopes, we solved the Lane-Emden equation numerically. After finding the
solution corresponding to a desired value of $q$ for a given value of
the polytropic index $n$, we sampled the resulting structure at
initial radii $r_0(i) \propto i^l$ to generate the $i$'th Lagrangian
element. The sampling parameter $l$ was varied to capture the full
range of mass, initial density and initial radius, and to give the
fastest convergence for studies in which the number of zones was
varied. When investigating the highest-velocity ejecta, we were
motivated by spherical truncation of self-similar acceleration (see
the Appendix) to distribute the Lagrangian zones evenly in the
parameter $x_0^{1/3}$. Runs were performed with between $500$ and
$10,000$ zones, depending on the convergence properties and the
fidelity desired for a particular result. 

Our calculations were performed with a maximum Courant number of
$0.5$, and with a coefficient of the quadratic artificial viscosity
law $c_0^2=2.5$. These values were chosen to optimize the energy and
entropy conservation, as well as the numerical shock structure. Small
spikes in the ejected mantle of the BSG (seen in figure
\ref{fig:pressurefigBSG}) are due to features (deviations from
smoothness) in the low-resolution progenitor that we interpolated for
our simulations; these are not caused by under-resolution of the
hydrodynamics. Energy was conserved to within $3\%$ in all our runs,
with the discrepancy growing during shock propagation and holding
steady after breakout. All calculations were carried out until the
thermal pressure was three orders of magnitude smaller than the
kinetic energy density in every zone; at this stage, pressure
gradients could safely be ignored. To determine the forward shock
velocity we compared the entropy distribution with the initial density
distribution. This comparison was made before the passage of a reverse
shock, for zones that experienced one. A slight loss of entropy
conservation (decreasing with increasing resolution) was observed in
the range of $10-100$ breakout times, and we were careful to record
the shock-deposited entropy of all zones by the time of breakout.

\subsection{Comparisons against simulations of realistic progenitors}

We have compared in detail both the pressure-based and the
harmonic-mean models for the final density distribution $\rho_f t^3$
against the output of the hydrodynamic code for the case of two
progenitor distributions: a $15M_\odot$ red supergiant progenitor
(Woosley \& Weaver 1995, courtesy of S. E. Woosley), and a $16M_\odot$
progenitor for supernova 1987A (Shigeyama \& Nomoto 1990, courtesy of
K. Nomoto). The comparison is made graphically in Figures
\ref{fig:pressurefigRSG}, \ref{fig:pressurefigBSG},
\ref{fig:harmfigRSG} and \ref{fig:harmfigBSG}, and the results are
summarized in Table \ref{t:errors}.

Both models are typically more accurate in the hydrogen envelope of
the red supergiant than that of the blue supergiant. This is to be
expected, because the parameters of these models were tuned to match
the results of simulation using polytropic analogs for convective
stars (\S \ref{SS:harmRSGs}). However, the effect of the
superadiabatic gradient on the high-velocity ejecta of the RSG is not
included in these polytropic analogs. The pressure-based model, which
includes more physics, better reflects the initial distribution of
this material and only incurs $\sim 34\%$ error (weighted by
$\log(v_f)$) as a result. The harmonic-mean model, on the other hand,
differs from simulation by about a factor of $\sim 2.5$ with the same
measure. The difference can also be seen in the figures.  This should
be considered an error of the polytropic analog, or a demonstration of
the effect of the superadiabatic gradient on the high-velocity ejecta.

In the blue supergiant as well, variations of the effective polytropic
index dominate the (log velocity-weighted) errors in the high-velocity
ejecta. An inspection of the progenitor reveals that $n_{\rm eff}$
approaches and then exceeds $3$ in the outer half of the radius, and
then becomes approximately constant in the outermost $10\%$. The
simulation and the models both attain the desired self-similar power
law index at the highest velocities. The deviations away from this
index are due to both the effects of sphericity (\S
\ref{SS:commonform}) and the variation in polytropic index: steeper
progenitors (higher $n_{\rm eff}$) give shallower ejecta (lower
$l_{\rho\,2}$) in self-similar theory (e.g., \cite{CS89}). The
pressure-based model does become shallower for lower $n_{\rm eff}$
because of the change in shock velocity, but not as shallow as the
true numerical behavior. As a result, an error accumulates because the
shape of the ejecta is slightly misrepresented, especially for
material that originated in the outer quarter or so of the radius.

It would be possible to expand the analytical description to include
variations in $n_{\rm eff}$ from their ideal values of $3/2$ and $3$
for red and blue supergiants. Such an analysis would be necessary for
the association of observational features in the high-velocity ejecta
with initial features of the progenitor. 

The mantle distribution of the red supergiant is badly represented in
both models. The pressure-based model suffers from a gross underestimation
of the entropy in this region, because we do not account for the
effect of the reverse shock. The harmonic-mean model does not cover
the RSG mantle at all, because it was fit only to the outer ejecta in
analogous polytropic simulations. 

For the blue supergiant, the effect of the reverse shock is not as
severe, and the pressure-based model fares quite well. Because it is
accurate in predicting the density of the mantle ejecta, the velocity
of the hydrogen-helium boundary is also accurately predicted. The
harmonic-mean model is aided by the weakness of the reverse shock as
well, but this model includes no information about the difference
between mantle and outer envelope. As a result, it incurs a
significant error in the BSG mantle material.

Note that the pressure-based model correctly predicts the density jump
(eq. [\ref{eq:rhojump}]) between the hydrogen and helium for both red
and blue supergiants. The effect of the reverse shock is apparent
inward in mass from this jump; it depresses the central density of the
red supergiant ejecta, producing an offset density maximum. We will
discuss this phenomenon further in the next section. 

Besides the stellar models presented so far, we also investigated the
explosion of model 3H11 of Nomoto, Iwamoto \& Suzuki 1995, provided to
us by Ken'ichi Nomoto. This model, meant to describe the progenitor
structure of SN193J, has an H envelope of mass $0.11M_\odot$, very
small compared to the $3.4M_\odot$ ejected. The envelope's density
structure is non-polytropic; indeed, there is a density minimum at the
base of the envelope and a local maximum within it (seen in figure 6
of Nomoto, Iwamoto \& Suzuki 195). The density then declines towards
the surface, but shallows out again, with $d\log\rho/d\log{\rm depth}$
decreasing from $1$ to $0.2$.  The reverse shock into the He mantle
stalls after it has traversed a mass comparable to the envelope mass,
leaving behind a jump in density. The shallow initial density law
leads to steeply declining final density in the outermost hydrogen
ejecta, as predicted by equation (\ref{eq:lrho2lp2}) and depicted by
Suzuki \& Nomoto (1995). The density jump and the variations of outer
power law were not reproduced by our pressure-based or harmonic-mean
models, although the helium mantle ejecta was fairly well represented.

\section{Conclusions} 

In this paper we develop analytical approximations that capture the
essential elements of spherically symmetric, adiabatic explosions in
media of finite mass and extremely stratified density distributions:
the progenitors of supernovae. Analytical predictions are motivated
and then calibrated by comparison to numerical simulations for
realistic red and blue supergiant progenitors. In separate sections we
address the velocity law of supernova shock waves (\S \ref{S:vshock}),
the expansion of the highest-velocity ejecta (\S \ref{S:highv}), and
two types of models for the final distribution of ejecta. One of these
models (\S \ref{S:pressuremodels}) specifies the pressure distribution
in the ejecta, and derives the final density distribution; the other
(\S \ref{SS:harmRSGs}) specifies the density directly as a function of
final velocity. The benefits and drawbacks of these models, in
application to red and blue supergiants, are summarized in Table
\ref{t:models}.

We emphasize the structure of the outer (typically hydrogen) envelope
ejecta in particular, because this layer is separated from the
underlying mantle material by a large drop in initial density. This
drop in density leads to an acceleration of the forward shock, followed
by a reverse shock into the mantle material as the forward shock
decelerates. Although our models can account for the density jump in
the ejecta caused by the acceleration of the forward shock (\S
\ref{SS:rhojump}), we make no attempt to model reverse shock
propagation. So, our models do not describe the mantle material
accurately, although this drawback is much less severe for blue
supergiants than for red ones.

No doubt the most useful conclusion of this work is that the ejecta
distributions from red supergiants are described by the single
parameter $q\equiv 1-M_{\rm env}/M_\star$. This simplification occurs
because these progenitors follow a sequence in parameter space (the
dashed line in Figure \ref{fig:paramfig}) as they lose mass,
regardless of the mechanism of stellar mass loss. When these stars
explode, their ejected hydrogen envelopes assume a distribution that
can be fit reliably by the simple, harmonic-mean models of \S
\ref{SS:harmRSGs}. The sequence of progenitors produces a sequence of
ejecta distributions (with fits tabulated in Table
\ref{t:harmonics}). The simplicity of these harmonic-mean models makes
them ideal for a systematic analysis of type II supernova light curves
from RSGs. Such an analysis would be useful in calibrating the
Expanding Photospheres Method, as discussed by Eastman, Schmidt and
Kirshner (1996).

However, this work is limited by the assumption of strict spherical
symmetry in the analytical models and in the numerical simulations
against which they are compared. The oblateness of a rotating
progenitor (Chevalier \& Soker 1989), the instability of accelerating
shock waves (albeit slowly growing; Chevalier 1990, Luo \& Chevalier
1994) and the Rayleigh-Taylor instability (e.g., Herant \& Woosley
1994) have interesting effects that are not accounted for
here. Moreover, by approximating supernovae as adiabatic,
non-gravitating point explosions we have neglected physics that is
important for the inner mantle material (where reverse shocks also
limit our accuracy).

Our model for the shock velocity (eq. [\ref{eq:vshock}]) is the
simplest possible formula that combines powers of the local variables
$m$, $r_0$ and $\rho_0$ in the progenitor and the blastwave energy
$E_{\rm in}$ in a manner consistent with both spherical and planar
scaling laws, without introducing dimensional scales. The success
(Figure \ref{fig:vshock}) of this formula implies that blastwave
shocks are controlled primarily by these scalings, even in situations
where the postshock flow is quite complicated, including, at times,
reverse shocks. This is very convenient, because it allows an accurate
estimation of the entropy distribution deposited by this forward shock
(although the entropy from a reverse shock is not accounted for). An
accurate prediction of the shock velocity and its dependence on the
stellar structure also permits a systematic analysis of the dynamics
of the high-velocity ejecta (\S \ref{S:highv}), including the maximum
velocity achieved by the ejecta (\S \ref{SS:vlimit}) and an analysis
of the burst of radiation that occurs at shock breakout (\S
\ref{SSS:emergence}). 

We find in \S \ref{SSS:relativity} that a star must be sufficiently
compact in order to produce any relativistic ejecta; for such stars,
we make a rough estimate of the rest mass of relativistic
ejecta. Applying these considerations to the supernova 1998bw and the
gamma-ray burst, GRB 9804025, that may be associated with it (Soffita
et al. 1998), we find that an explosion of relatively small ejected
masses ($1M_\odot$) and large energy ($\gtrsim 10^{52}$ ergs) could
provide enough energy in relativistic ejecta both for the gamma-ray
burst (Galama et al. 1998), and to power the inferred relativistically
expanding synchrotron shell (Kulkarni et al. 1998). Woosley et
al. (1998) considered models for SN1998bw with larger ejected masses
in order to fit the supernova light curve, and found accordingly
smaller energies in the relativistic ejecta. Even assuming the
existence of this ejecta, liberating its energy in a brief period
requires an unusually high circumstellar density, as Woosley et
al. point out.

The velocity at which stellar material expands into space is related
to the velocity of the shock front that set it into motion. However,
the nature of this relationship is very different for material deep in
the ejected envelope than for material near the stellar surface.
Simple considerations (\S \ref{S:highv}) reveal that the
highest-velocity (initially outermost) material is cast away at some
multiple of the shock velocity that traversed it, although the value
of this factor varies with initial depth. At lower velocities, the
smoothness of the pressure distribution controls the density
distribution of the ejecta from deep within the progenitor. Here the
shock velocity determines the final velocity indirectly, by means of
the entropy distribution. The transition between these two behaviors
is the result of a complicated, spherical rarefaction flow. But,
simple models for the pressure distribution (\S
\ref{S:pressuremodels}) reproduce this transition quite accurately,
especially in polytropic progenitors for which the high-velocity
behavior is well understood (\S \ref{SS:commonform}). Features in the
ejecta at low velocities, most notably the structure of the ejected
mantle, can be understood on the basis of approximate pressure
equilibrium.

One consequence of pressure equilibrium for the low-velocity ejecta is
the presence a sharp density jump at the base of the outer envelope:
the helium ejecta is denser and has a lower entropy than the hydrogen
ejecta that surrounds it. This jump, quantified in \S
\ref{SS:rhojump}, is greater in the ejecta from red supergiants (about
a factor of $30$) than from blue supergiants (about a factor of
three). It results directly from the density difference between
envelopes in a progenitor, and the shock acceleration that occurs
there. As we have discussed, this feature is unlikely to be affected
by the development of reverse shocks or the Rayleigh-Taylor
instability, and may have interesting observational implications in
the epoch of recombination.

The much larger drop in density from helium to hydrogen between red
and blue supergiants also leads to a much stronger reverse shock in
RSGs than in BSGs. This limits the accuracy of our pressure-based
models, which do not include the entropy deposited in the helium
mantles (particularly in RSG mantles) by the reverse shock. It also
has important consequences for the velocity of radioactive material in
the ejecta. Consider, for instance, the fate of the innermost
$0.5M_\odot$ of ejecta in the $15M_\odot$ RSG and the $16M_\odot$ BSG
progenitors depicted in Figures \ref{fig:pressurefigRSG} and \ref{fig:pressurefigBSG}. In the numerical
simulations, which include reverse shocks, this half solar mass is one
to three times more dense at a given time in the RSG ejecta than in
the BSG ejecta (the comparison is straightforward, since the ejected
masses are quite similar). Because the outer envelope harbors almost
all of the energy in the final state, its expansion velocity is
essentially fixed. So, the expansion velocity of the mantle is
determined by its density relative to the outer envelope, which
depends on the difference in entropy between the two regions. This, in
turn, is affected by the reverse shock.  Our pressure-based models
indicate what might happen if the reverse shock never developed, and
the mantle were decelerated by a compression wave instead: it would
then lack the reverse shock entropy, and be much denser (and expanding
slower) in comparison to the outer envelope. Although these concepts
are the products of spherically-symmetric models, similar
considerations should apply in realistic explosions as well.

It is noteworthy that the reverse shock develops inward (in mass) from
the density jump between the helium and hydrogen ejecta, so that there
is a density peak at the outer helium ejecta in red supergiants
(Figure \ref{fig:pressurefigRSG}). 

In BSGs the reverse shock is much less severe than for RSGs, and (at
least for the model $16 M_\odot$ BSG provided by Nomoto) seems to
develop far inward in mass from the hydrogen-helium boundary. The
pressure-based model tracks the mantle ejecta quite well. Indeed, the
difference between the mantle and the outer envelope can be ignored
entirely without too much error: such is the spirit of the Chevalier
\& Soker (1989) model for SN1987A, whose elements we have incorporated
into our harmonic-mean model for BSGs. Here, the mantle and the inner
region of the hydrogen ejecta can be grouped together as a single
region of gently declining density. 

Regardless of the mantle, it appears that the ejecta density
distribution in the bulk of the hydrogen envelope is a flat or gently
declining function of velocity (or mass) for both red and blue
supergiants. At some break velocity, the structure of $\rho_f(v_f)$
rolls over toward the steeply declining power law it must assume at
high velocities. For red supergiants, a quantitative statement can be
made: density falls by only a factor of $20-40\%$ in the inner half of
the hydrogen ejecta mass. (This density drop decreases with increasing
mass ratio $q$, despite the fact that the inner power law $l_1$ gets
steeper [Table \ref{t:harmonics}], but note that the hydrogen envelope
is a smaller fraction of the ejecta as $q$ increases.) 

Chevalier (1976), making an analogy with Sedov blastwaves of different
indices, suggested that a detached shell (density maximum) should form
in the ejecta at shells for which $d\ln\rho_0/d\ln r_0$ decreases
smoothly through the value $-2.53$ in the progenitor. If this were
strictly true, a detached shell would form in the outer envelope
ejecta of all the progenitors we have considered. No such density
maxima actually formed in the outer envelopes, although density maxima
did occur in the ejected mantles (of RSGs especially, as described
above). The presence of density maxima in the ejected mantles depends
on the strength of the reverse shock, and does not appear to be
related to Chevalier's proposed mechanism.

It is clear from our simulations that when a progenitor deviates from
a polytropic structure in the outer half of its radius, perhaps
because of inefficient convection or the details of radiative
transfer, the resulting high-velocity ejecta distribution bears the
imprint of this effect. This is apparent when the ejecta from
simulations with realistic stellar progenitors are compared against
the ejecta of polytropic models. Our analysis of the high-velocity
ejecta of polytropic models is limited in applicability by this
effect. On the other hand, the relative constancy of the shock
velocity parameter $\beta_1$, the postshock acceleration ratio
$(v_f/v_s)_{\rm p}$, and the truncation radius $R_t$ for planar
acceleration all suggest that it may be possible to extend our
semi-analytical theory to accommodate variations in $n_{\rm eff}$. 
With such a theory, it may even be possible to constrain models
of inefficient convection with the early light curves of type II
supernovae.

Our pressure-based model, being derived from a simple model for the
entropy deposited in the blastwave shock, is a \emph{local} theory
that relates the final density $\rho_f$ of a mass shell with its
initial density $\rho_0$, initial radius $r_0$, and enclosed ejected
mass $m$. A strong reverse shock, whose passage cannot be predicted by
these local variables alone, invalidates this model in the mantles of
stars. But, the local theory is exceptionally accurate where it is
valid. 

Our harmonic-mean models, on the other hand, are \emph{global} models
for the ejecta distributions of supernovae, in the sense that they
specify the general run of final density versus velocity that results
from a specific type of progenitor structure. These models are
appealing for their simplicity, although they contain little
information about the dynamics of envelope expulsion in supernovae. 

\acknowledgements The authors are indebted to Stan Woosley and
Ken'ichi Nomoto for providing supernova progenitor structures, and to
J. Kelly Truelove for providing the computer code that we modified for
the purposes of this paper. CDM thanks John Monnier, Ben Metcalf and
George McLean for many valuable suggestions, Lars Bildsten for
clarifying stellar structure and evolution and reviewing a draft of
this work, and Douglas Leonard for discussing light curves. We also
wish to thank our referee, Roger Chevalier, for insightful and helpful
comments. CDM was supported in part by a National Science Foundation
Graduate Fellowship and an ARCS fellowship. The research of both CDM
and CFM is supported in part by the National Science Foundation
through NSF grant AST 95-30480.

\appendix
\section{Fully Lagrangian, self-similar treatment of shock emergence}
\label{S:apdx} 

We present a calculation of the self-similar, planar flow in a power
law density distribution with an edge, the problem posed by Gandel'man
\& Frank-Kamenetsky (1956) and solved by Sakurai (1960). Unlike
previous treatments, we use a Lagrangian approach for the pre-breakout
flow as well as the post-breakout flow. This has the advantage that
variables can be compared against their postshock values, rather than
against their breakout values (which previously were related to
postshock values with an Eulerian treatment). To do this, we first
define the space and time coordinates $x \equiv 1-r/R_\star$ and
$\Delta t\equiv t-t({\rm breakout})$. The time at which a mass shell
experiences the shock is $\Delta t_0(m)$. For our self-similar
coordinate we choose $\eta\equiv \Delta t/\Delta t_0(m)$, and for the
flow variable we choose $S(\eta)=x(m)/x_0(m)$. Our variables $\eta$
and $S(\eta)$ are both unity in the initial state, and decrease toward
$-\infty$ as the flow progresses. Breakout occurs when $\Delta t=\eta=0$.

To maintain consistency with previous authors, we define the initial
density distribution as $\rho_0(m)=k_1 [R_\star x_0(m)]^n$, and the shock
velocity to be $v_s = -k_2 [R_\star x_0(m)]^{-\lambda}$, where $\lambda\equiv
n\beta_1$. Then, $\Delta t_0 = -(R_\star x_0)^{1-\lambda}/k_2(1+\lambda)$. 

Considering the time variable $\Delta t$ to be parameterized by $\eta$ and
$x_0$, the constraint of constant time, $d\Delta t=0$, gives
$d\eta=-\eta(1-\lambda)dx_0/x_0$. This relation allows us to take the
necessary derivatives to solve for the flow. First of these is the
density, defined in planar geometry by: 
\begin{equation} 
\rho(\eta,x_0) = \rho_0\left(\frac{dx_0}{dx}\right)_{\Delta t} = 
\frac{\rho_0}{S(\eta)-(\lambda+1)\eta S'(\eta)}. 
\end{equation}
Now, we evaluate the pressure as $p(\eta,x_0)= p_s(\rho/
\rho_s)^\gamma$, where $p_s$ and $\rho_s$ are the postshock pressure
and density. The acceleration in planar geometry is,
$R_\star \ddot{x}=-(dp/dr)_{\Delta t}/\rho$ $=-(dp/dr_0)_{\Delta t}/\rho_0$. From the
definitions of $S(\eta)$ and $\eta$, on the other hand, $\ddot{x}=x_0
S''(\eta)/t_0^2$. Combining the two expressions for the acceleration,
we have the following differential equation for $S(\eta)$ to describe planar,
self-similar flow: 
\begin{equation}\label{eq:selfsimeqn}
(1+\lambda)^2 S'' = -\frac{2}{\gamma+1} 
\left(\frac{\gamma-1}{\gamma+1}\right)^\gamma 
\left\{\frac{n-2\lambda}{[S-(\lambda+1)\eta S']^\gamma}
-\gamma (\lambda-1) \eta
\frac{\lambda S'+(\lambda-1)\eta S''}
{[S-(\lambda+1)\eta S']^{\gamma+1}}\right\}. 
\end{equation}

The initial condition for this equation is at the time $\eta=1$, when
a mass shell is hit by the shock wave. Because the mass shell is in
its initial position at this time, and because its postshock density
and velocity are given by the Rankine-Hugoniot jump conditions for
strong shocks, 
\begin{equation} \label{eq:selfsiminit}
S(1)=1,\,\,\,\,\,\,\,\, S'(1) = \frac{2}{(\gamma+1)(\lambda+1)}. 
\end{equation}
Note that the ratio $v/v_s$ at any point in the flow is equal to
$(\lambda+1)S'(\eta)$, for the value of $\eta$ appropriate to the time
and mass shell considered. 

We used the technique described by Sakurai to determine $\lambda$ (or
equivalently $\beta_1$) for $\gamma=4/3$ and $n=3/2$ and 3. Then we
solved for the evolution of a mass shell's position using equation
(\ref{eq:selfsimeqn}), starting with the initial conditions of
equation (\ref{eq:selfsiminit}). We continued the numerical solution
out to $|\eta|>10^8$, and used the known asymptotic solution to find
the limit $S'(\eta\rightarrow -\infty)$. For the purpose of
studying the solution, we then fit the resulting velocity history
$v/v_s$ in powers of $(2-\eta)^{-1/3}$. This parameterization was
chosen to show the asymptotic dependence of $S(\eta)$ as
$\eta\rightarrow -\infty$, and also to be finite for all $\eta<1$. We
obtain,
\begin{eqnarray}
\frac{v(m,\eta)}{v_s(m)} &\simeq& 2.1649-\frac{1.0656}{(2-\eta)^{1/3}}
+\frac{0.5872}{(2-\eta)^{2/3}} -\frac{1.1235}{2-\eta}
\,\,\,\,\,\,\,\,(n = 3/2), \nonumber \\
\frac{v(m,\eta)}{v_s(m)} &\simeq& 2.0351-\frac{0.7638}{(2-\eta)^{1/3}}
+\frac{0.2535}{(2-\eta)^{2/3}} -\frac{0.7861}{2-\eta}
\,\,\,\,\,\,\,\,(n = 3).
\end{eqnarray}
These fits, meant to capture the late-time evolution accurately, are 
accurate to within $2\%$ and $1.5\%$, respectively, for $\eta<-1$,
although they miss the breakout value ($v_f/v_s=6/7$) by $34\%$ and
$14\%$, respectively.  For the purpose of studying the
truncation of the self-similar acceleration law, as discussed in
\S\ref{S:highv}, it is useful to make similar fits based on the
normalized distance coordinate $S(\eta)=x/x_0$ rather than the time
coordinate $\eta$:
\begin{eqnarray} \label{eq:vfvsfromS}
\frac{v(m,S)}{v_s(m)} &\simeq&
2.1649-\frac{1.2504}{(2-S)^{1/3}}+\frac{0.8473}{(2-S)^{2/3}}-
\frac{0.9817}{2-S}
\,\,\,\,\,\,\,\,(n = 3/2), \nonumber \\
\frac{v(m,S)}{v_s(m)} &\simeq& 2.0351
-\frac{0.8260}{(2-S)^{1/3}}+\frac{0.3094}{(2-S)^{2/3}}- 
\frac{0.5244}{2-S}
\,\,\,\,\,\,\,\,(n = 3).
\end{eqnarray}
Much like the previous fits, these are both valid to within $1\%$ for $S<-1$,
but they miss the breakout value by $9\%$ and $16\%$, respectively. 

Litvinova \& Nad\"ezhin (1990) suggested that spherical effects would
act to truncate the planar acceleration once a shell had reached a
truncation radius $R_t\gtrsim R_\star$.  Using our solution to
the self-similar expansion, we can evaluate this prescription for
different truncation radii and compare it to the results of fully
spherical calculations (\S \ref{SS:commonform}). The normalized
position $S_t$ corresponding to $R_t$ is, 
\begin{equation} \label{eq:Stx0}
S_t = -\frac{R_t/R_\star-1}{x_0}
\end{equation}
Taking $R_t = 3R_\star$ as an illustrative example (Figure
\ref{fig:vfvsfig}), the self-similar solution can be fit as a polynomial
in $x_0^{1/3}$ in accordance with equation (\ref{eq:Stx0}): 
\begin{eqnarray}\label{eq:vfvstrunc}
\frac{v_f(r_0)}{v_s(r_0)} &\simeq&
\left(\frac{v_f}{v_s}\right)_p \left(1 
-0.4522  x_0^{1/3}
+0.1717   x_0^{2/3}
-0.04328 x_0\right) 
 \nonumber \,\,\,\,\,\,(n=3/2), \\
\frac{v_f(r_0)}{v_s(r_0)} &\simeq&
\left(\frac{v_f}{v_s}\right)_p \left(1 
-0.3174 x_0^{1/3}
+0.0373 x_0^{2/3}
+0.0199 x_0\right)
\,\,\,\,\,\, (n=3).  
\end{eqnarray}
These fit the self-similar solution quite well, with maximum errors
$0.3\%$ and $0.5\%$ for $n=3/2$ and $3$, respectively, in the entire
range $0<x_0<1$. Clearly these are quite similar to the outcome of
full spherical dynamics as given by equation
(\ref{eq:vfvsUniversal}).

\newpage

\figcaption[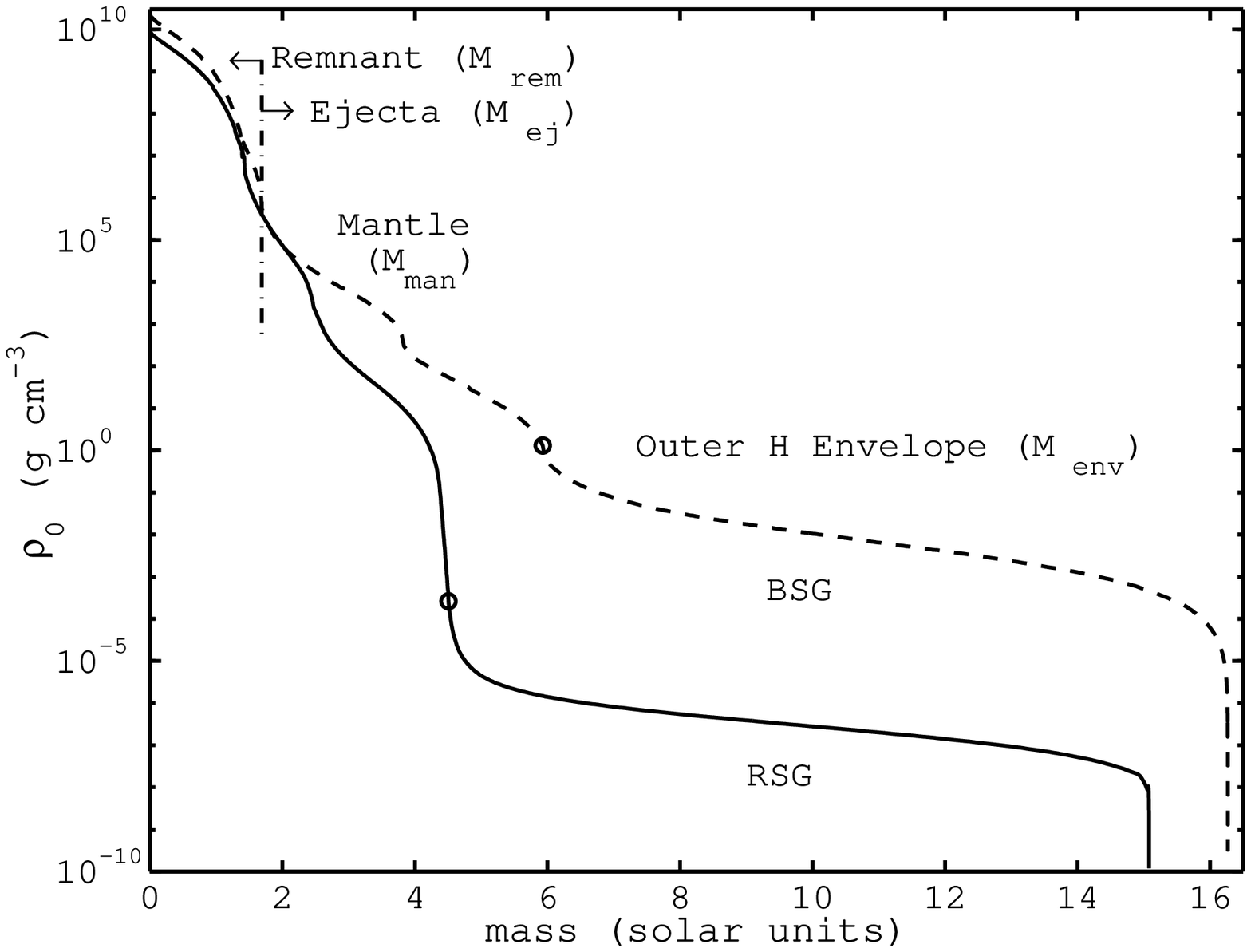]{The red and blue supergiant progenitors used
in this paper. The $15 M_\odot$ RSG (\emph{solid line}, Woosley \&
Weaver 1995) was provided by Stan Woosley. The $16 M_\odot$ BSG
(\emph{dashed line}, Shigeyama \& Nomoto 1990) was provided by
Ken'ichi Nomoto. The boundaries between remnant and ejecta
(\emph{dash-dot line}), and between mantle and outer envelope material
($M_\alpha$, \emph{circles}), are marked in both cases. We take the remnant to be
the edge of the iron core; this gives remnant masses of $1.48$ and
$1.69 M_\odot$ for the RSG and BSG,
respectively. \label{fig:progenfig}}

\figcaption[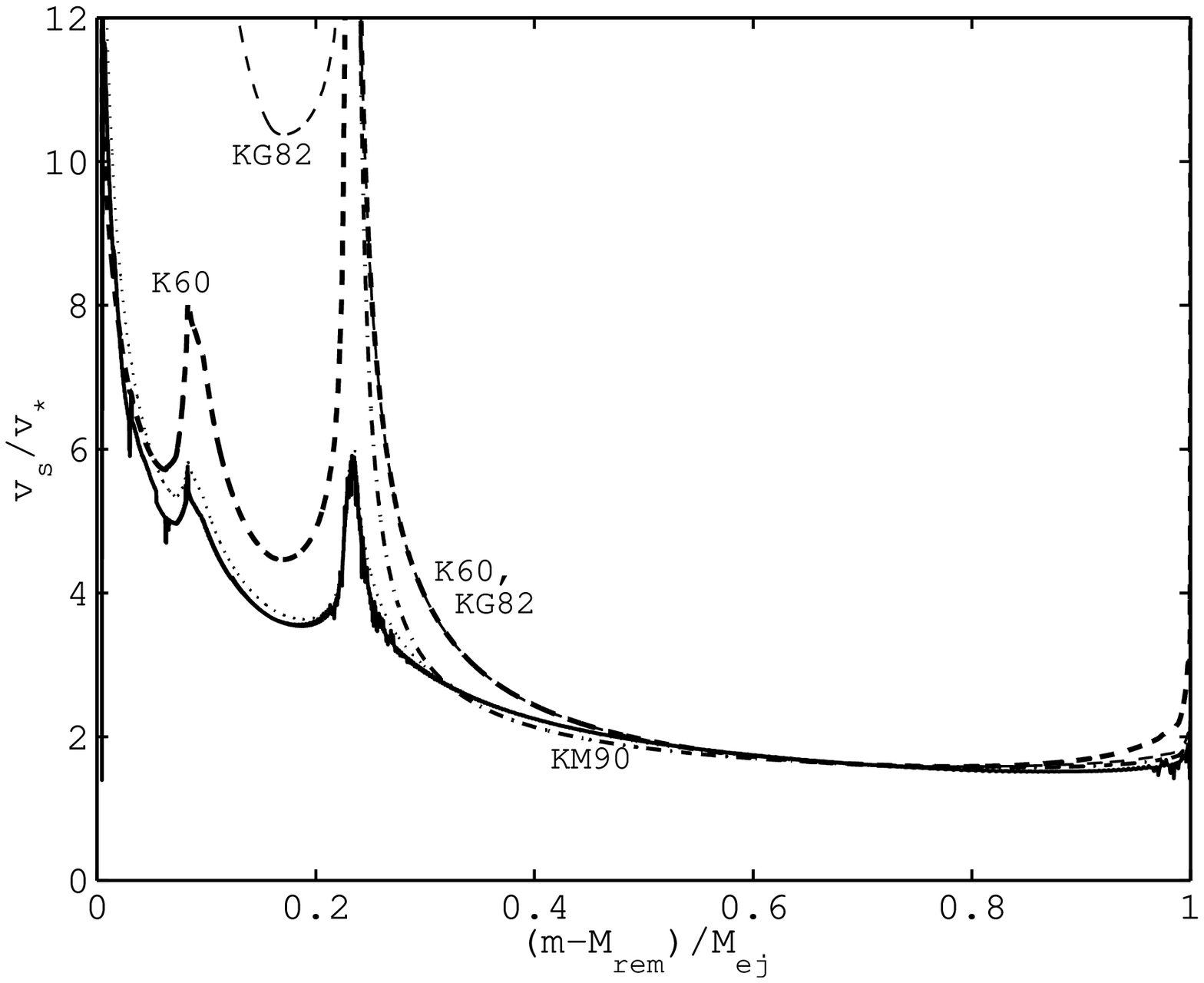]{The forward shock velocity in a numerical
simulation of the $15 M_\odot$ RSG progenitor S15A of Woosley \& Weaver
1995 (\emph{solid line}), and a comparison between our shock velocity formula
(eq. [\ref{eq:vshock}], \emph{dotted line}) and the formulae of
Kompaneets (1960, \emph{thick dashed line}), Klimishin \& Gnatyk (1982,
\emph{thin dashed line}), and Koo \& McKee (1990, \emph{dash-dot 
line}). Except for our formula, shock velocities have been normalized
to match the numerical value at $\m=0.7$. The constant $R_{s,t}$ in
the model of Koo \& McKee has been chosen to optimize the shock
velocity in the outer envelope. To be fair, ours is the only formula
designed specifically for the context of supernovae.\label{fig:vshock}}

\figcaption[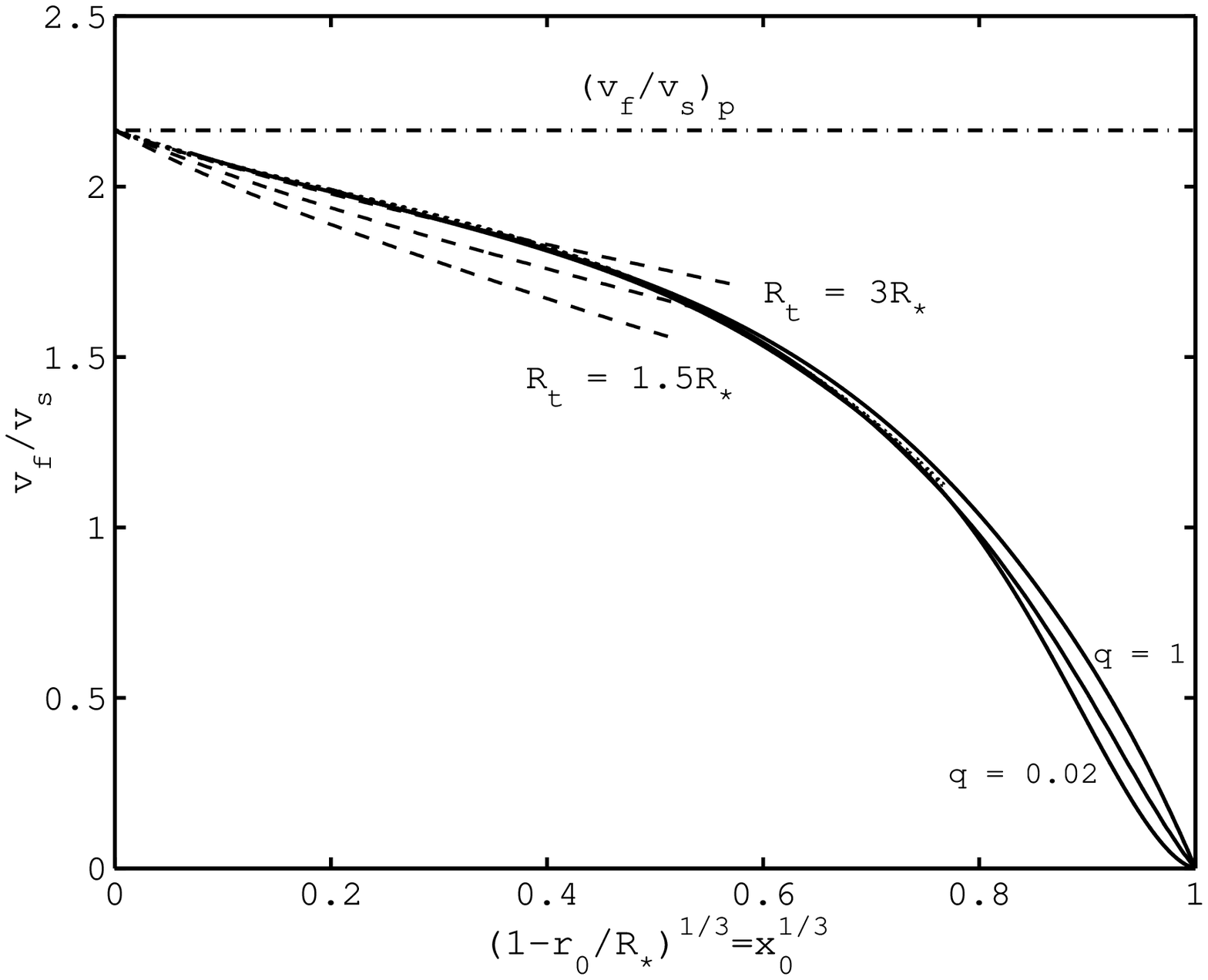]{The ratio of shock to final velocity for a
sequence of simulated explosions in polytropes without mantles (index
$n=3/2$, $q = 0.02$, $0.3$ and $1$), parameterized by the radial
variable $x_0^{1/3}$. If the progenitors were homologous copies of one
another, dimensionless quantities such as this would not vary. The
progenitors are most similar at the outermost radii, a feature that
holds for this ratio as well. The ratio in each simulation approaches
the planar, self-similar value (\emph{dash-dot line}). The suggestion
of Litvinova \& Nad\"ezhin (1990) that spherical expansion may
truncate self-similar expansion at a radius $R_t$ gives a definite
prediction of the variation of this ratio with depth (\emph{dashed
lines}, for $R_t=1.5$, $2$ and $3R_\star$). Our fit to the dependence
of $v_f/v_s$ with $x_0$ is also plotted (eq. [\ref{eq:vfvsUniversal}],
\emph{dotted line}).
\label{fig:vfvsfig}}

\figcaption[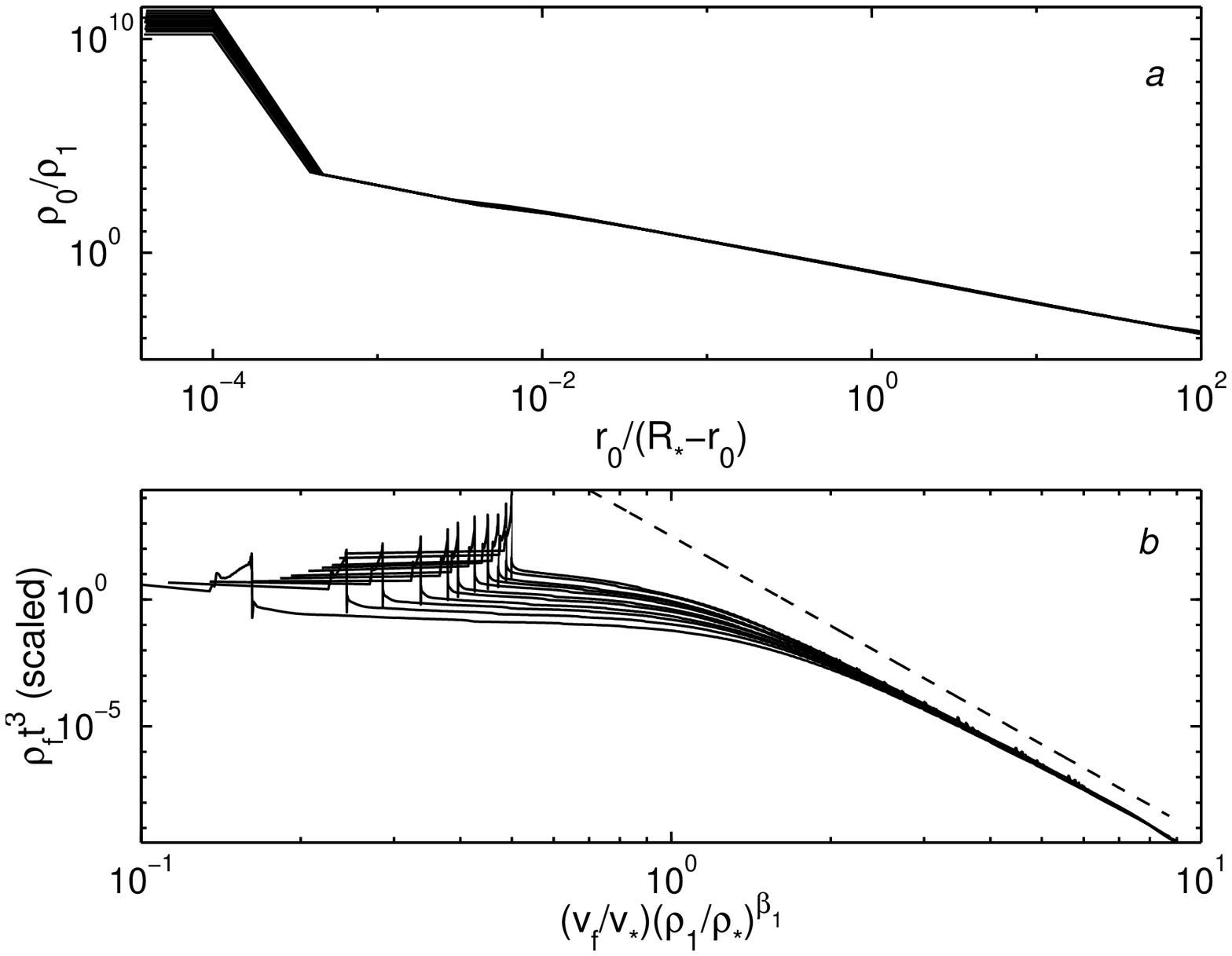]{Similarity of the outer regions of
polytropic envelopes is reproduced as similarity of the high-velocity
ejecta. \emph{a}: The sequence of polytropic progenitors used to
simulate red supergiants (\S \ref{S:RSGs}), plotted as normalized
initial density $\rho_0 / \rho_1$ against the variable
$r_0/(R_\star-r_0)$, in which $\rho_0 / \rho_1$ is a power law for the
region in which polytropic envelopes share a common form
(\S\ref{SS:rho01}). \emph{b}: The ejecta distribution produced in
these models: normalized final density against normalized final
velocity. For comparison, the self-similar planar power law is also
shown (\emph{dashed line}). \label{fig:commonfig}}

\figcaption[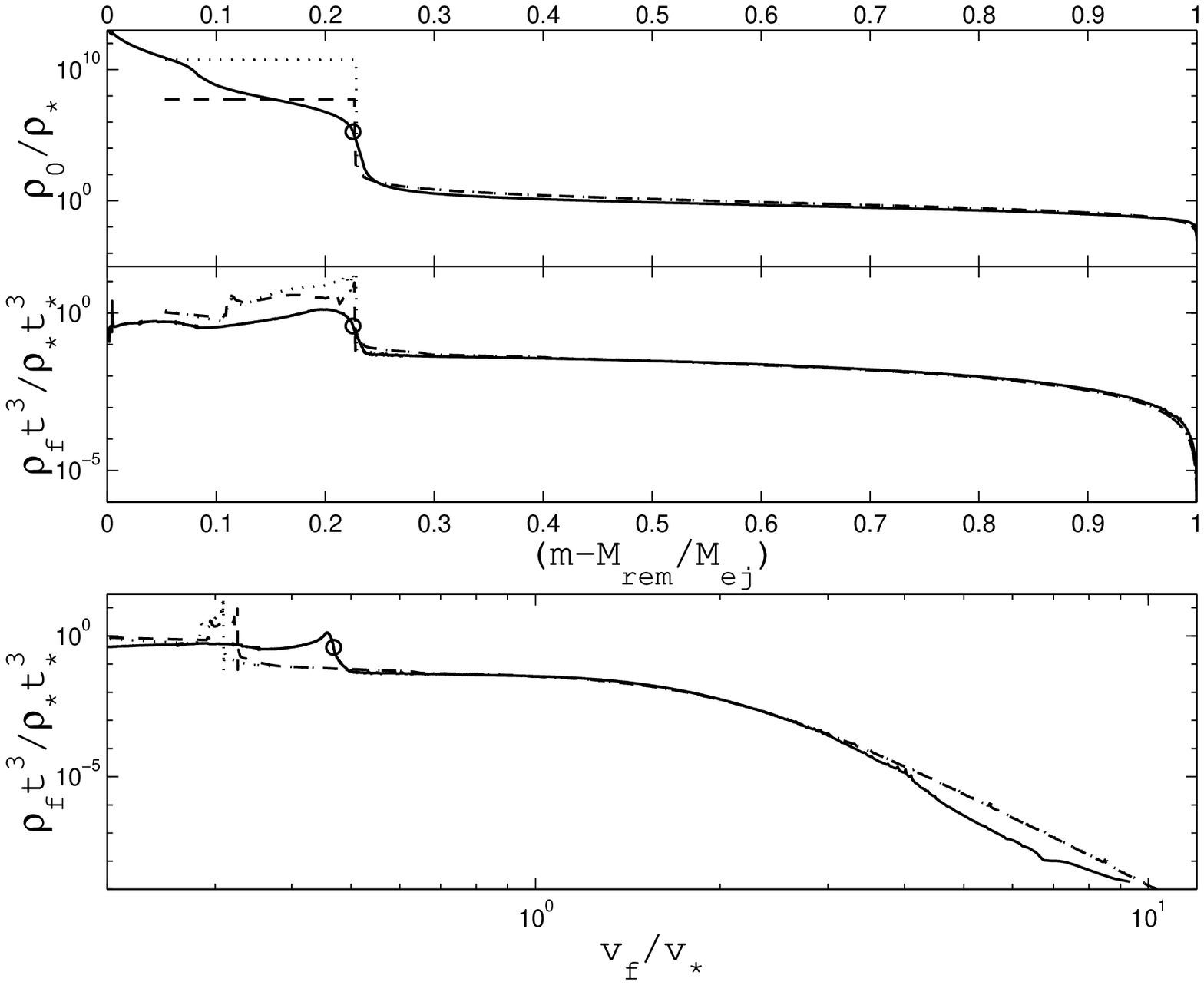]{This figure demonstrates the applicability
of polytropic models to red supergiant envelopes, as well as the
insensitivity of the outer envelope ejecta to the size and structure
of the mantle. The red supergiant model of Woosley \& Weaver (1995,
\emph{solid lines}) has a mantle radius $R_{\rm man}\simeq 0.02
R_\star$ and a complicated mantle structure resembling $\rho_0\propto
r^{-3}$. The polytropic models have mantles of constant density, with
$R_{\rm man} =10^{-3}R_\star$ (\emph{dashed lines}) and $R_{\rm
man}=10^{-4}R_\star$ (\emph{dotted lines}). Nevertheless, the outer
envelope ejecta is well-modeled by both progenitors. The
superadiabatic gradient is significant in the subsurface layer of the
progenitor, which produces a high-velocity density distribution that
differs from polytropic models. \label{fig:polyrsgfig}}

\figcaption[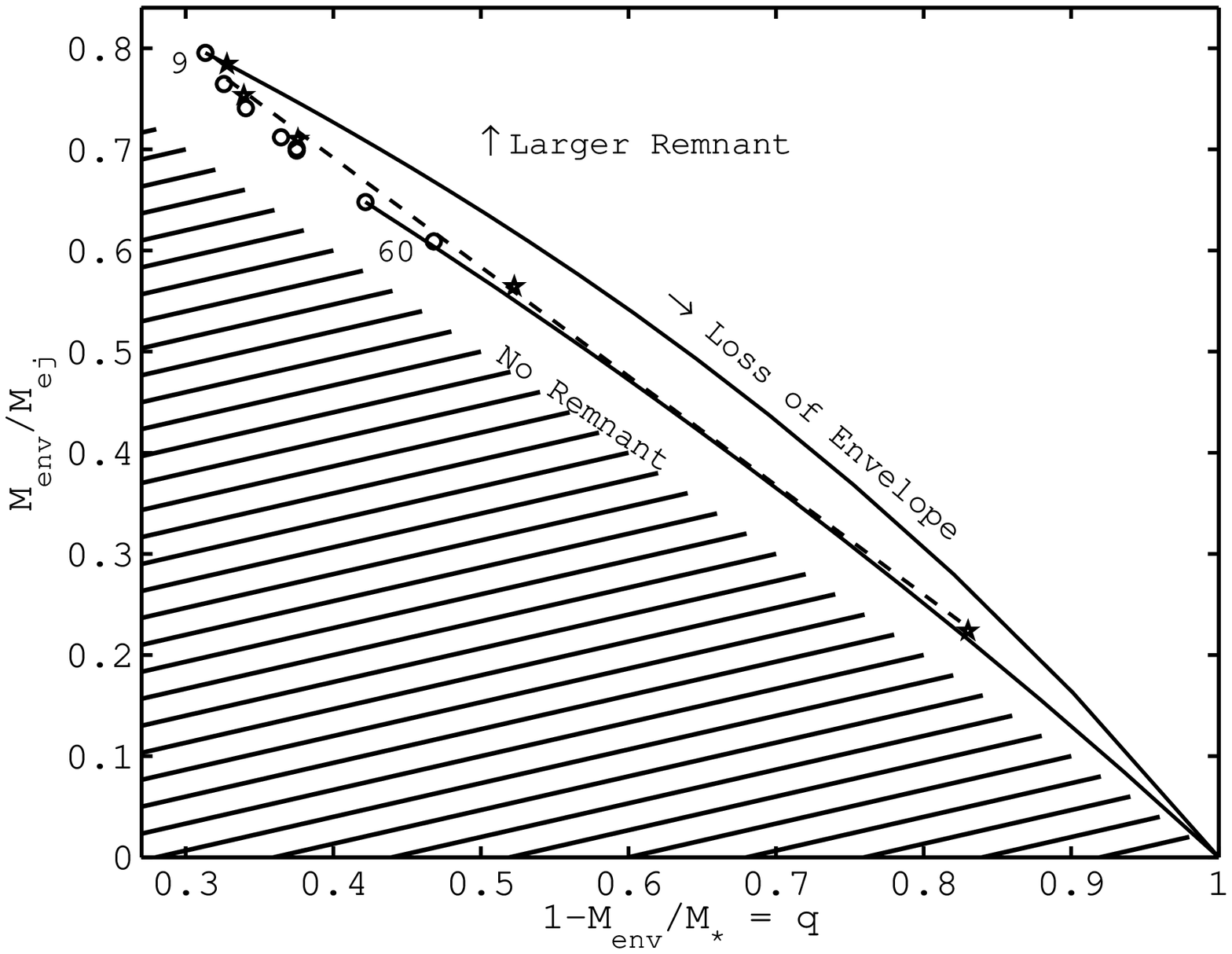]{The mass ratios that parameterize type II
supernovae. Plotted is the envelope fraction in the ejecta, versus the
ratio $q$ of core to stellar mass. If remnant masses are correctly
estimated, progenitors must lie in the swath between the solid lines
regardless of mass loss. As a star loses its envelope, it travels from
the point corresponding to no mass loss (\emph{circles}, from Maeder
1992, for 9, 12, 20, 25, 40 and 60 $M_\odot$) along a curve
(\emph{solid lines}, for $9$ and $40 M_\odot$) towards the point
$(1,0)$. The results of lone stellar mass loss at solar metallicity
(\emph{stars}, Woosley \& Weaver 1995) for 9, 12, 20, and 25 $M_\odot$
lie in the interior.  Stars initially above about $27 M_\odot$ will
only retain their envelopes if the metallicity is low.  The remnant
mass cannot be negative (\emph{hatched region}), but it could be zero
in a thermonuclear explosion. \emph{Dashed line}: the sequence of
models (eq. [\ref{eq:qline}]) considered in \S
\ref{S:RSGs}. \label{fig:paramfig}}

\figcaption[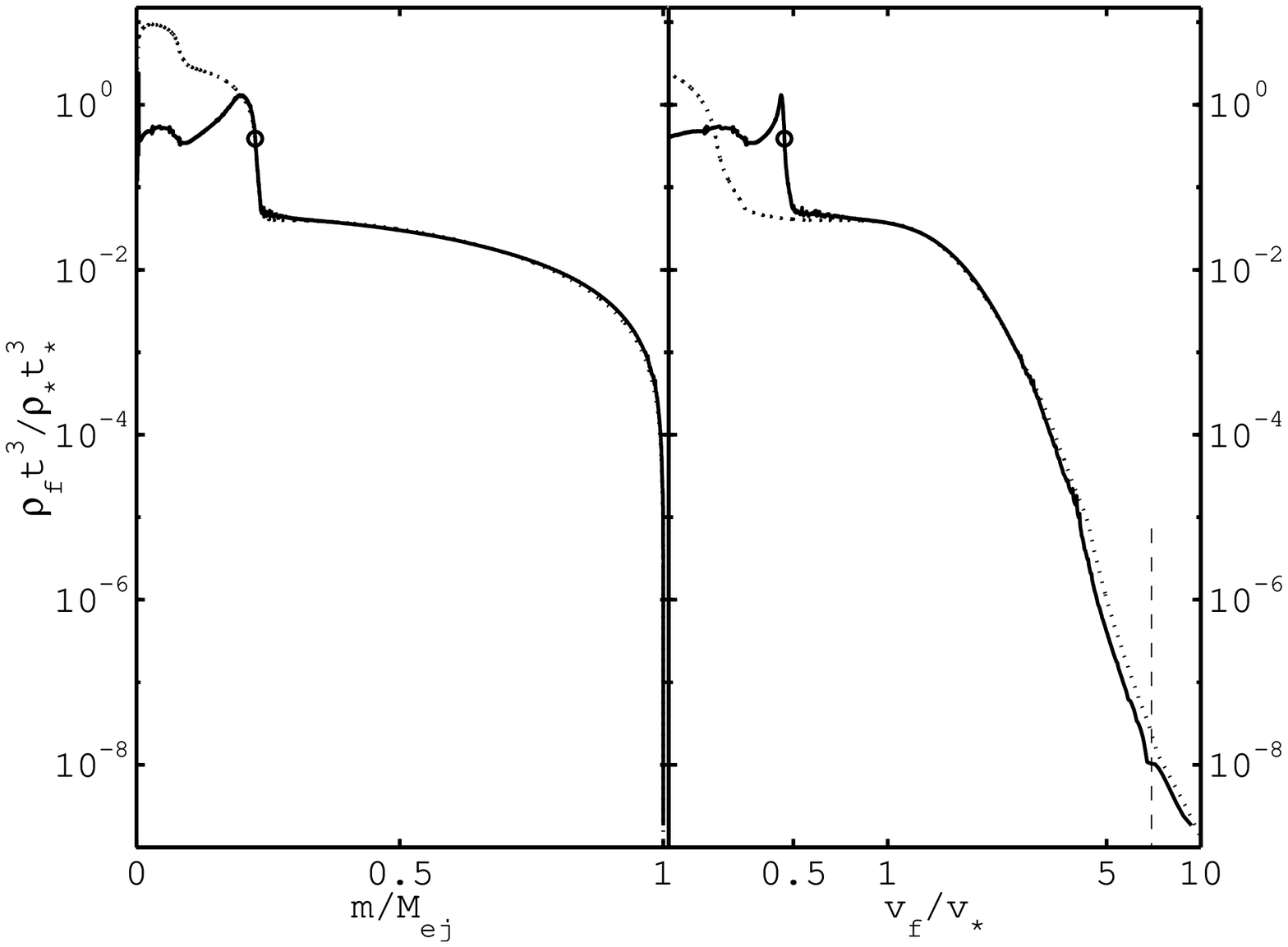]{A comparison of the pressure-based
model (\S \ref{S:pressuremodels}, \emph{dotted lines}) with the result
of simulation for realistic progenitors (\emph{solid lines}, with
\emph{circles} at the H-He boundary). The maximum ejecta velocity is
marked (\S \ref{SS:vlimit}, \emph{dashed lines}).  Left side:
normalized final density $\rho_f t^3/\rho_\star t_\star^3$ with
respect to the normalized ejecta mass $\m \equiv m/M_{\rm ej}$. Right side:
$\rho_f t^3/\rho_\star t_\star^3$ versus the normalized velocity
$v_f/v_\star$. The $15M_\odot$ red supergiant progenitor model of
Woosley \& Weaver (1995) was the input into both the simulation and
the analytical model. The parameter $\alpha$ was chosen according to
the formulae in Tables \ref{t:harmonics} and \ref{t:radharms}.  Note
that the models fail interior to the base of the hydrogen envelope,
because of the entropy deposited in the reverse shock, but the jump in
density at the base of the outer envelope is correctly estimated (\S
\ref{SS:rhojump}). 
\label{fig:pressurefigRSG}}

\figcaption[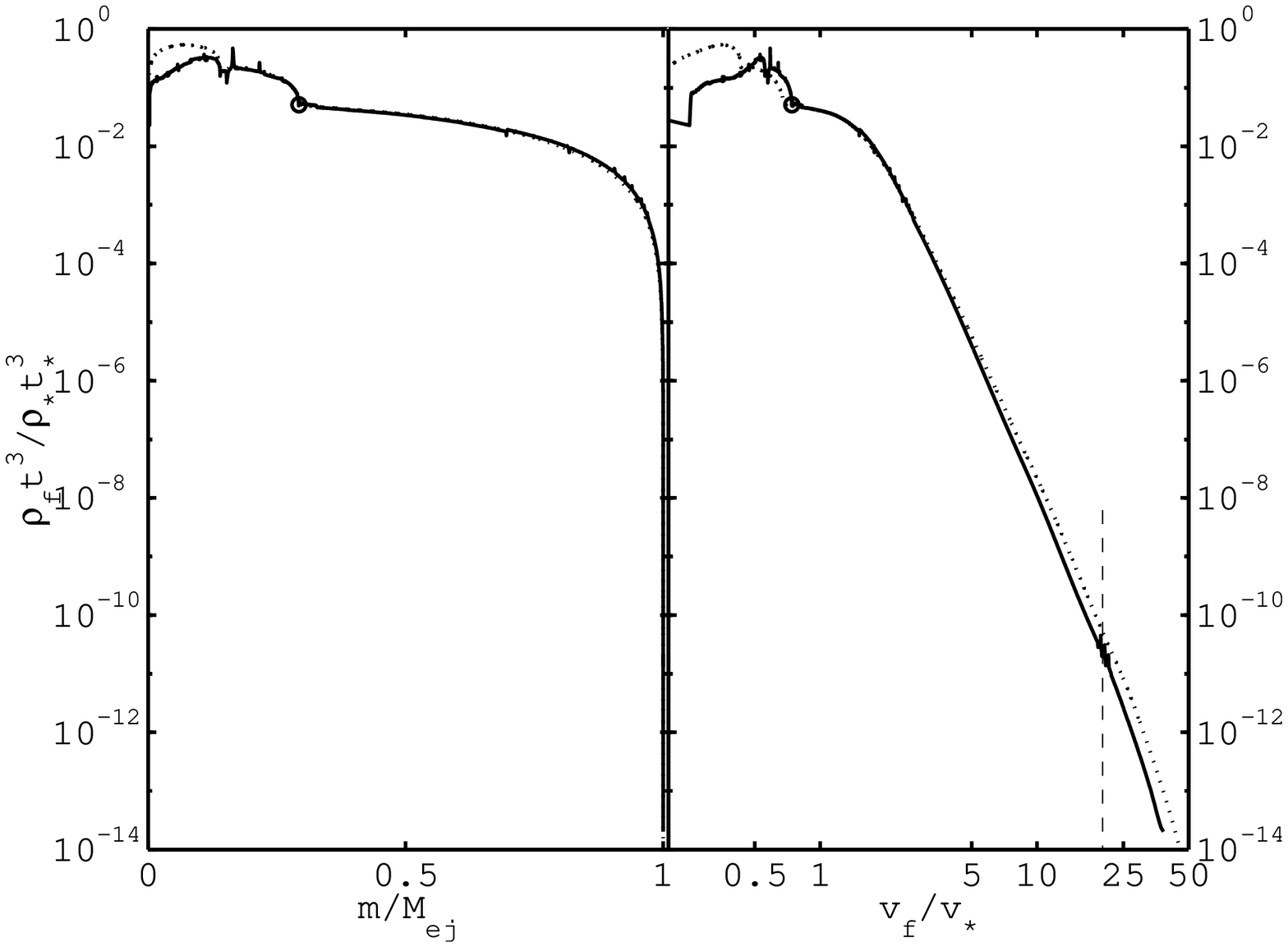]{A comparison of the pressure-based
model with simulation, but here the progenitor was taken to be the $16
M_\odot$ blue supergiant model of Shigeyama \& Nomoto (1990). As in
Figure \ref{fig:pressurefigRSG}, the \emph{solid line} is the
simulation, with \emph{circles} at the He-H boundary; the \emph{dotted
line} is the output of the pressure-based model, and a \emph{dashed
line} marks the maximum velocity to which the adiabatic approximation
is valid, which is approximately the speed limit for the ejecta. Note
the lesser effect of the reverse shock for the BSG compared to the
RSG, the higher maximum velocity, and the greater range of ejecta
densities. Again, the jump in density at the H-He boundary is
correctly determined by the model.
\label{fig:pressurefigBSG}}

\figcaption[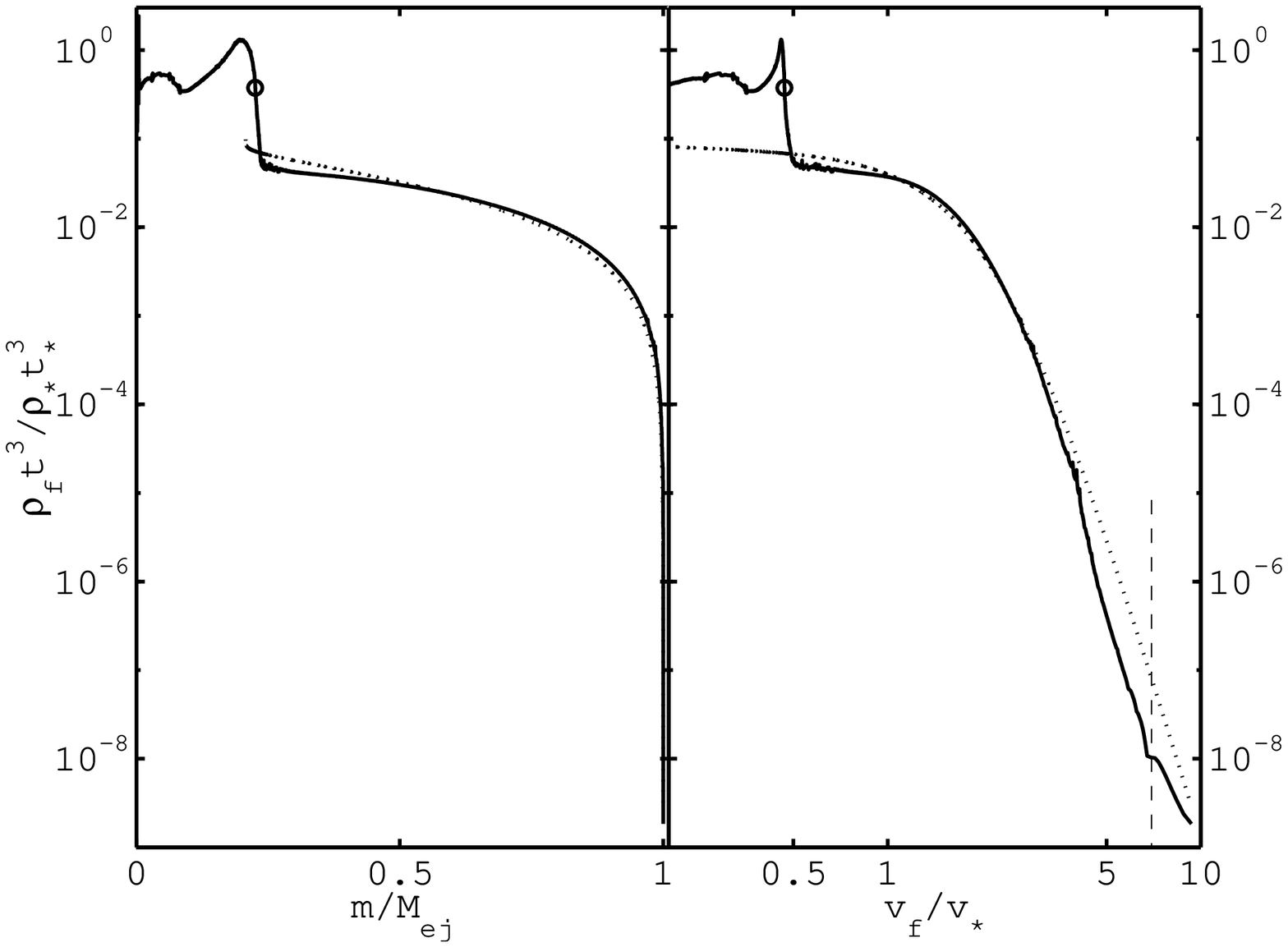]{ Application of the harmonic-mean model (\S
\ref{SS:harmRSGs} and Table \ref{t:harmonics}: \emph{dotted lines}) to
numerical simulation (\emph{solid lines}, with \emph{circles} at the
H-He boundary), for the RSG progenitor of Woosley \& Weaver
(1995). Again, the \emph{dashed line} shows the maximum velocity of
the ejecta.  For RSG progenitors, the model is derived from a
polytropic progenitor, as tabulated in Table \ref{t:harmonics}. It is
meant to describe the hydrogen envelope only. Since the models
prescribe $\rho_f(v_f)$, the model mass coordinate is obtained by
integrating inward from $\m =1$ at high velocities.
\label{fig:harmfigRSG}}

\figcaption[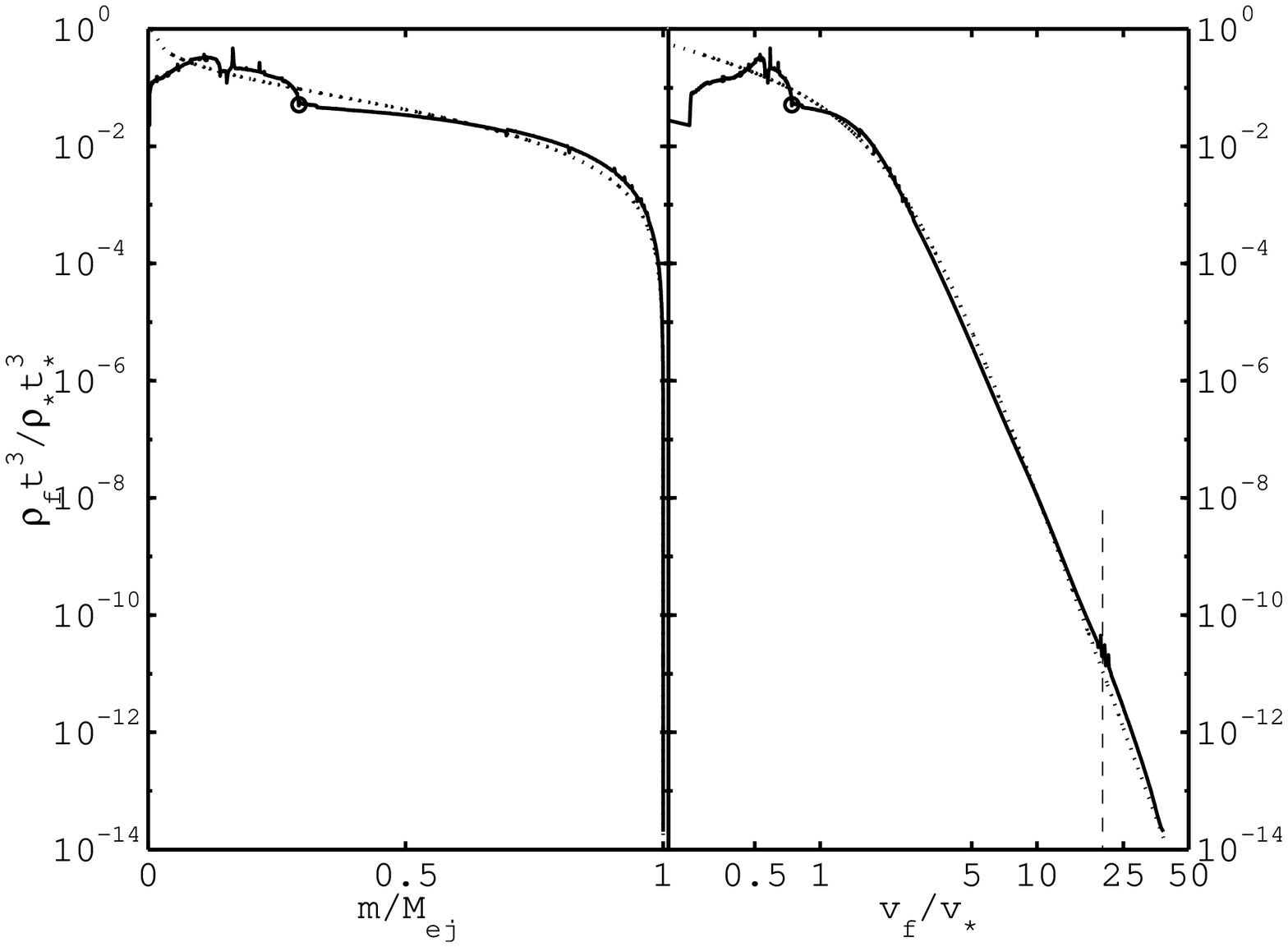]{A comparison of the harmonic-mean model (\S
\ref{SS:harmRSGs}: \emph{dotted lines}) with
simulation (\emph{solid lines}), here for the blue supergiant
progenitor of Shigeyama \& Nomoto (1990). For the BSG progenitor, the
constraints of mass and energy conservation are imposed, and the inner
power-law is chosen to agree with the self-similar rarefaction model
of Chevalier \& Soker 1989 (Table \ref{t:radharms}). This model is
therefore meant to cover the entire ejecta distribution.
\label{fig:harmfigBSG}}

\newpage
\begin{deluxetable}{lllcc}
\tablecaption{Models for the ejecta density.
\label{t:models}}
\tablehead{%
\colhead{} &\colhead{Pressure-Based Models (\S
\ref{S:pressuremodels})} 
	&\colhead{Harmonic-Mean Models (\S \ref{SS:harmRSGs})} }
\startdata
Model output & $\rho_f(m)t^3$	& $\rho_f(v)t^3$  \nl
Type of model& Local, Lagrangian & Global, Eulerian \nl
Flexibility  & Responds to detailed stellar structure & Fixed
analytical form  \nl
Benefits     & Gives jump in $\rho_f$ at mantle & Simple \nl
Drawbacks    & Requires initial structure & Insensitive to detailed structure\nl
RSG models\tablenotemark{\dagger}   & Parameter $\alpha$ fit as $\alpha(q)$
	  & 3 parameters 	fit with $q$ \nl
  & H envelope most accurate &H envelope only \nl
BSG, Ib, Ic models\tablenotemark{\ddagger}   & Approximate $\alpha=0$
   & Generalized model of CS89\tablenotemark{*}\nl 
 & Applies to mantle & Approximates mantle \nl
\enddata
\tablenotetext{\dagger}{Models for red supergiants are based upon polytropic
models for the progenitors; these are primarily determined by one
parameter, $q$.} 
\tablenotetext{\ddagger}{Radiative progenitors are not
easily modeled with simple progenitors, but this is counterbalanced by
the fact that features in their ejecta are less pronounced.}
\tablenotetext{*}{CS89: \cite{CS89}.}
\end{deluxetable}

\begin{deluxetable}{llllc}
\tablecaption{Quantities associated with self-similar, planar expansion.
\label{t:selfsim}}
\tablehead{%
\colhead{Quantity} &\colhead{$n=3/2$}&\colhead{$n=3$} & \colhead{Comments}
}
\startdata
$\lambda$& 0.2863 	& 0.5573 	& Method of Sakurai (1960) 	\nl
$\beta_1$& 0.1909 	& 0.1858 	& Method of Sakurai (1960) 	\nl
	 & 0.2071 	& 0.2071 	& Appx. of Whitham (1960) 	\nl
$\left(\frac{v_f}{v_s}\right)_{\rm p}$ 
	 & 2.1649 	& 2.0351	& Derived in Appendix\nl
$w_{\rho_0}$ & 3/5 	& 3/4		& Initial density distribution	\nl
$w_{v_s}$& -0.1145 	& -0.1394	& Equation (\ref{eq:ws})	\nl
$w_{s}$  &-0.3218	&-0.3965	& Equation (\ref{eq:ws})	\nl
$w_\rho$ &1.3436	&1.4181		& Equation (\ref{eq:ws})	\nl
$w_p$	 &1.3624	&1.3624		& Equation (\ref{eq:ws})	\nl
%
%
\enddata

\end{deluxetable}

\begin{deluxetable}{llllc}
\tablecaption{Formulae for high-velocity distributions of ejecta
quantities. 
\label{t:highv}}
\tablehead{%
\colhead{Quantity} &\colhead{High-velocity formula}
&\colhead{Validity} 
}
\startdata
\sidehead{Convective outer envelopes ($n = 3/2$):}
 $v_f/v_\star$ & $2.06  (\rho_1/\rho_\star)^{-0.08}(1-\m)^{ -0.1145}
(1+0.30\dmhat)$ 
& $\dmhat\lesssim0.8$	\nl
 $\rho_f t^3/\rho_\star t_\star^3$ & $ 0.079  (\rho_1/\rho_\star)^{-0.23}
(1-\m)^{1.344} (1 + 0.67\dmhat)$ & $\dmhat\lesssim 0.8$	\nl
 $p_f t^4/ p_\star t_\star^4  $ & $ 0.0027  (\rho_1/\rho_\star)^{0.02}
(1-\m)^{1.362}(1 + 0.96\dmhat)$
&$\dmhat\lesssim 0.7$ 	\nl
\tablevspace{.05 in}\sidehead{Radiative outer envelopes ($n=3$):}
 $v_f/v_\star$& $ 1.92 (\rho_1/\rho_\star)^{-0.05} (1-\m)^{ -0.1394}
 (1 - 0.23\dmhat -0.095\dmhat^2)$ & $\dmhat\lesssim 0.9$ \nl
 $\rho_f t^3/\rho_\star t_\star^3$  & $ 0.081 (\rho_1/\rho_\star)^{0.14}
(1-\m)^{1.418} (1 + 0.49\dmhat + 0.61\dmhat^2)$ & $\dmhat\lesssim 0.9$	\nl
$p_f t^4/ p_\star t_\star^4  $ & $ 0.0026(\rho_1/\rho_\star)^{0.01}
(1-\m)^{1.362}(1 + 1.03\dmhat + 0.37\dmhat^2)$
&$\dmhat\lesssim 0.75$ 	\nl
%
%
\enddata

\tablecomments{ Final distributions depend on the parameter
 $\rho_1/\rho_\star$ defined by equation (\ref{eq:outerform}),
 describing the subsurface layers of the progenitor (\S
 \ref{SS:rho01}). Characteristic values such as $v_\star$ are
 combinations of the ejecta mass, energy, and radius (\S
 \ref{S:scaling}). For each quantity, the distribution limits to a
 power law at the highest velocities. The factor that depends on
 $\dmhat$ describes the deviation from this power law toward lower
 velocities, primarily due to the onset of spherical expansion. Here,
 $\dmhat\equiv [(1-\m)(\rho_1/\rho_\star)^{-1}]^{1/3(n+1)}$. These
 formulae are valid in the region for which envelopes of the same
 value of $n$, but different internal structures, share a common form
 in numerical simulations. Significant digits are apportioned
 according to accuracy: most accurately determined are the power law
 exponents, i.e., the exponents on $(1-\m)$.}
\end{deluxetable}

\begin{deluxetable}{llllc}
\tablecaption{Parameters for modeling the H ejecta from RSGs. 
\label{t:harmonics}}
\tablehead{%
\colhead{Parameter} &\colhead{Value} 
}
\startdata
\sidehead{Typical RSG sequence (eq. [\ref{eq:qline}]):}
$\alpha$&$5.58q^2-0.10q-1.17$\nl
\tablevspace{.02 in}
$\rho_{\rm break}t^3/\rho_\star t_\star^3$ 
	& $10^{-3}(5.21q^2-10.12q+5.40)$\nl
$v_{\rho\,{\rm break}}/v_\star$ & $0.93q + 1.94 $ \nl
$l_{\rho\,1}$	&$-3.06q^2+2.03q-0.46$\nl
$l_{\rho\,2}$	&$-11.731$\nl
$y_\rho$&$4.5$\nl
$p_{\rm break}t^4/p_\star t_\star^4$& 
	$10^{-4}(-9.87q^2+4.67q+4.87)$\nl
$v_{p\,{\rm break}}/v_\star$ & $0.80q+1.85$  \nl
$l_{p\,1}$	&$1.68q-1.89$\nl
$l_{p\,2}$ 	& $-11.895$\nl
$y_p$	&$3.8$\nl
\tablevspace{.05 in}\sidehead{Partial derivatives, for extrapolation:}
$\partial\alpha/\partial(M_{\rm env}/M_{\rm ej})$ 
	& $ -17.9q +3.67$\nl
\tablevspace{.02 in}
$\partial(\rho_{\rm break}t^3/\rho_\star t_\star^3)/
	\partial(M_{\rm env}/M_{\rm ej})$ 
	&$ 10^{-3}(-10.12q+9.53)$\nl
$\partial(v_{\rho\,{\rm break}}/v_\star)/
	\partial(M_{\rm env}/M_{\rm ej})$ 
	&$ 1.14q - 1.43$\nl
$\partial l_{\rho\,1}/
	\partial(M_{\rm env}/M_{\rm ej})$ 
	&$3.82q - 1.38$\nl
$\partial(p_{\rm break}t^4/p_\star t_\star^4)/
	\partial(M_{\rm env}/M_{\rm ej})$ 
	&$ 10^{-3}(1.53q-0.43)$\nl 
$\partial (v_{p\,{\rm break}}/v_\star)/
	\partial(M_{\rm env}/M_{\rm ej})$ 
	& $0.28 q -0.81$ \nl
$\partial l_{p\,1}/
	\partial(M_{\rm env}/M_{\rm ej})$ 
	&$-0.74q-2.13$\nl
%
%
\enddata

\tablecomments{The parameter $\alpha$ of equation (\ref{eq:rhomodel}),
and the parameters used in equations (\ref{eq:harmrho}) and
(\ref{eq:harmp}). The latter specify the harmonic-mean approximation
for the hydrogen ejecta of red supergiants described in \S
\ref{SS:harmRSGs}; in the former, $\alpha$ specifies the more
complicated pressure-based model of \S
\ref{S:pressuremodels}. Parameters that depend on $q$ are derived from
the one-parameter family of progenitors described by equation
(\ref{eq:qline}) and shown as the dashed line in Figure
\ref{fig:paramfig}. Partial derivatives corresponding to changing the
remnant mass without changing $M_\star$ or $M_{\rm env}$ (i.e., at
fixed $q$) are derived from comparison with a second sequence of
models offset by $0.05$ in $M_{\rm env}/M_{\rm ej}$.  The fit for $\alpha$
is valid in the range $0.3\leq q\leq 0.6$; others are valid for
$0.3\leq q\leq 0.8$.}
\end{deluxetable}

\begin{deluxetable}{llllc}
\tablecaption{Fiducial ejecta models for radiative progenitors. 
\label{t:radharms}}
\tablehead{%
\colhead{Parameter} &\colhead{Value} &\colhead{Comments}
}
\startdata
$\alpha$	&$0$		& Typical value\nl
$\rho_{\rm break}t^3/\rho_\star t_\star^3$ 
		& $1.91\times 10^{-3}$	
				& Energy and mass conservation\nl
$v_{\rho\,{\rm break}}/v_\star$ 
		& $2.30$ 	& Energy and mass conservation\nl
$l_{\rho\,1}$	& $-1.06$ 	& Theory of Chevalier \& Soker 1989\nl
$l_{\rho\,2}$	&$-10.176$	& Self-similar value ($n=3$) \nl
$y_\rho$	&$\sim 4.5$	& From RSG models\nl
$p_{\rm break}t^4/p_\star t_\star^4$
		&$\sim 2.4\times 10^{-4}$ & Gleaned from simulation\nl
$v_{p\,{\rm break}}/v_\star$ 	
		& $\sim 2.3$	& From RSG models  \nl
$l_{p\,1}$	&$2.22$		& Theory of Chevalier \& Soker 1989\nl
$l_{p\,2}$ 	& $-9.77$	& Self-similar value ($n=3$)\nl
$y_p$		&$3.8$		& From RSG models\nl
\enddata

\tablecomments{Parameters for modeling the entire ejecta of radiative
stars. Without a sequence of reliable models to fit, these models are
only educated guesses. The harmonic-mean model (all but $\alpha$) is a
combination of the self-similar rarefaction hypothesis of Chevalier \&
Soker 1989 with self-similar planar expansion for $n=3$, and softening
parameters $y_{\rho}$, $y_p$ taken from the RSG fits. We simply assume
$\alpha=0$ to specify a pressure-based model. The pressure-based model
reflects the detailed progenitor structure through the entropy
deposited in the shock, whereas the harmonic-mean model is
inflexible. Note that the effect of the reverse shock is less severe
for these progenitors than for RSGs, so the mantle is better represented.}
\end{deluxetable}

\begin{deluxetable}{lllll}
\tablecaption{Calibration of ejecta models for realistic
progenitors. \label{t:errors}} \tablehead{%
\colhead{} & \multicolumn{2}{c}{RSG\tablenotemark{*}}
 & \multicolumn{2}{c}{BSG\tablenotemark{\dagger}}\\
\cline{2-3}\cline{4-5}\\
\colhead{Error estimate} &\colhead{H envelope} &\colhead{Mantle}
&\colhead{H envelope} &\colhead{Mantle}  }
\tablecolumns{5}
\startdata
\sidehead{Pressure-based model (\S \ref{S:pressuremodels})}
$\left<\delta\rho_f/\rho_f \right>_{\m}$ & $0.07$ &$6.1$&$0.10$&$0.71$ \nl
$[\delta\rho_f(\m)/\rho_f(\m)]_{\rm max}$ &$2.1$&$21$&$7.5$ &$5.8$ \nl
$\left<\delta\rho_f/\rho_f \right>_{\log (v_f)}$ &$0.34$ &$4.5$ &$0.96$&$3.0$ \nl
$[\delta\rho_f(v_f)/\rho_f(v_f)]_{\rm max}$ &$1.4$&$32$&$4.5$&$15$\nl

\tablevspace{.05 in}\sidehead{Harmonic-mean model (\S \ref{SS:harmRSGs})}
$\left<\delta\rho_f/\rho_f \right>_{\m}$ &$0.20$&\nodata & $0.31$ &$0.70$ \nl
$[\delta\rho_f(\m)/\rho_f(\m)]_{\rm max}$ & $4.5$ &\nodata &$0.83$ &$22$ \nl
$\left<\delta\rho_f/\rho_f \right>_{\log (v_f)}$ &$1.5$ &$5.0$ &$0.53$&$3.2$ \nl
$[\delta\rho_f(v_f)/\rho_f(v_f)]_{\rm max}$ &$10$ &$18$ &$2.4$ & $19$\nl
\enddata
\tablenotetext{*}{$15 M_\odot$ RSG of Woosley \& Weaver 1995, courtesy
of Stan Woosley} \tablenotetext{\dagger}{$16 M_\odot$ BSG of Shigeyama
\& Nomoto 1990, courtesy of Ken'ichi Nomoto.}  \tablecomments{Mean and
maximum relative errors in the density models of this paper, for both
the Lagrangian and Eulerian functions $\rho_f(m)t^3$ and
$\rho_f(v_f)t^3$. Mean errors are computed according to the formula
$\left<\delta\rho_f/\rho_f\right>_{x}\equiv
\exp\sqrt{\left<\left[\log(\rho_{\rm
model}(x)/\rho_f(x))\right]^2\right>_x}-1$, where $\left<\;\right>_x$
is a mean weighted evenly in the variable $x$. The maximum error is
defined consistently: an error of $1$ implies a factor of $2$ (in
either direction) between $\rho_f(x) t^3$ and $\rho_{\rm model}(x)
t^3$. In RSGs, the harmonic mean approximation does not cover the
mantle. Errors in the H envelope are dominated by variations in the
effective polytropic index from its ideal value ($3/2$ for RSGs and
$3$ for BSGs) in the outer layers of the progenitor. Errors in the
mantle are dominated by the effect of the reverse shock, which is not
addressed in these models.  }
\end{deluxetable}

\end{document}